\documentclass[acmsmall,screen,nonacm]{acmart}

\usepackage{array,tabularx,longtable}

\newcolumntype{Y}{>{\raggedright\arraybackslash}X}
\newcolumntype{L}[1]{>{\raggedright\arraybackslash}p{#1}}

\newcommand{\MWM}{\textsc{MWM}}
\newcommand{\MWPM}{\textsc{MWPM}}
\newcommand{\Edm}{\textsc{Edmonds}}
\newcommand{\Gab}{\textsc{Gabow}}
\newcommand{\Scal}{\textsc{Scaling}}

\newcounter{pseudoline}
\newenvironment{pseudocode}[2]{%
  \par\addvspace{0.9em}\noindent\begin{minipage}{\linewidth}
  \color{black}\normalfont\rmfamily
  \hrule height 0.8pt\vspace{0.3em}
  \noindent{\normalsize\textbf{#1}\enspace #2}\par\vspace{0.25em}
  \hrule height 0.4pt\vspace{0.35em}
  \begingroup\fontsize{9.5}{12}\selectfont
  \setlength{\parindent}{0pt}\setlength{\parskip}{0pt}
  \setcounter{pseudoline}{0}
}{%
  \endgroup\vspace{0.25em}\hrule height 0.5pt
  \end{minipage}\par\addvspace{0.7em}
}
\newcommand{\pseudometa}[2]{%
  \noindent\textbf{#1:}\enspace #2\par\vspace{0.15em}
}
\newcommand{\pline}[2][0]{%
  \stepcounter{pseudoline}%
  \noindent\makebox[1.8em][r]{\arabic{pseudoline}:}\hspace{0.6em}%
  \begin{minipage}[t]{\dimexpr\linewidth-2.4em\relax}
    \setlength{\leftskip}{#1em}\strut #2\strut
  \end{minipage}\par\vspace{1.5pt}
}

\begin{document}

\title[Experimental Evaluation of Maximum Weighted Matching Algorithms]{An Experimental Evaluation of Maximum Weighted Matching Algorithms}
\subtitle{Edmonds, Gabow, and Scaling on General Graphs}

\author{Abdulaziz Alfaraj}
\orcid{0009-0003-7714-0052}
\affiliation{%
  \institution{King Fahd University of Petroleum and Minerals}
  \city{Dhahran}
  \country{Saudi Arabia}}
\email{g202185070@kfupm.edu.sa}

\author{Ahmed Al-Herz}
\affiliation{%
  \institution{King Fahd University of Petroleum and Minerals}
  \city{Dhahran}
  \country{Saudi Arabia}}
\email{alherz@kfupm.edu.sa}

\begin{abstract}
Maximum weighted matching is a classic problem in combinatorial optimization. Edmonds gave the first polynomial time algorithm  for the problem on general graphs. The fastest algorithm is Gabow's algorithm which runs in $O(n(m+n\log{n}))$ for general weights and Duan, Pettie and Su algorithm with time complexity of $O(m\sqrt{n}\log{nN})$ using integer weights. We compare the three algorithms: \Edm{}, a blossom search organized as in Galil, Micali and Gabow (1986) with one lazily pruned queue of candidate edges; \Gab{}, the same search with the supporting forest and tree-blossom merging of Gabow (2018) for those candidates; and \Scal{}, the Hybrid scaling algorithm of Duan, Pettie and Su (2018), which solves the perfect-matching variant. To the best of our knowledge, \Gab{} and \Scal{} are the first implementations of Gabow's supporting-forest and tree-blossom-merging candidate structure and of the Hybrid algorithm. We use two sets of instances in our experiment, Suite~A and Suite~B. Suite~A is used to compare \Edm{} and \Gab{} on graphs obtained from SuiteSparse and randomly generated graphs with up to 55 million vertices, as well as on 24 structured graphs with about one million vertices that we built from dense bipartite or odd-cycle structures. Suite~B is used to compare all three algorithms on 50 randomly generated graphs that admit maximum-weight perfect matchings. \Edm{} and \Gab{} break ties by the same rule and performed the same search on every graph, so their differences stem from the data structure and its maintenance. \Gab{} was faster than \Edm{} on all 24 structured graphs, by 4 to 53\%, but slower on 74 of the other 75 graphs, by up to 39\%, and it used 1.2 to 2.0 times as much memory on every graph. \Scal{} was the fastest algorithm on all 29 Erd\H{o}s--R\'enyi graphs, 2.6 to 11.8 times faster than \Edm{}, and on bipartite random and community graphs, whereas \Edm{} was fastest on random geometric graphs, by a factor of 26 to 33, and on disconnected clique and biclique blocks. The largest overhead of \Gab{} is due to tearing down search trees; the slowdown of \Edm{} and \Gab{} on Erd\H{o}s--R\'enyi graphs generated with specific seeds is due to repeated contraction of large blossoms; and \Scal{} is the fastest algorithm on these graphs, though its slowdown on geometric graphs is due to longer searches and to dismantling deeply nested blossoms. Our implementations and experimental data are publicly available at \url{https://github.com/AzizFaraj/maximum-weight-matching}.
\end{abstract}



\maketitle

\section{Introduction}

Given a simple, undirected graph $G=(V,E)$ and a nonnegative edge weight function $w:E \to \mathbb{R}_{\geq 0}$, the Maximum Weighted Matching (\MWM{}) problem asks for a set of pairwise vertex-disjoint edges of greatest total weight. In this article, we study the problem on general graphs, which corresponds to Problem~4 in Galil's classification of matching problems~\cite{Galil1986Survey}. In non-bipartite graphs, an alternating search (alternating between matched and unmatched edges) may close an odd alternating cycle, called a blossom. Edmonds' idea was to contract the blossom, continue the search in the contracted graph, and later lift the matching through it~\cite{EdmondsWeighted1965,EdmondsFlowers1965}. 
Two later algorithms improve on the $O(mn\log n)$ bound of Galil, Micali and Gabow~\cite{GalilMicaliGabow1986}, where $m=|E|$ and $n=|V|$. Gabow's data structures find the next tight edge faster, giving $O(n(m+n\log n))$ time~\cite{Gabow1990,Gabow2018}, and the Hybrid algorithm of Duan, Pettie and Su processes integer weights up to $W$ one bit at a time, giving $O(m\sqrt n\log(nW))$~\cite{DuanPettieSu2018}. To our knowledge, the only implementations available are variants of Edmonds' algorithm, such as Blossom~IV~\cite{CookRohe1999}, Blossom~V~\cite{Kolmogorov2009}, Blossom~VI~\cite{ArkhipovKolmogorov2026}, LEMON~\cite{LEMON2011} and the implementation of Mehlhorn and Sch\"afer in LEDA~\cite{MehlhornSchaefer2002}. An experimental study is essential to answer the open question, posed by Mehlhorn and Sch\"afer, of whether sophisticated data structures used in theoretically faster but more complex algorithms (like Gabow's algorithm and the scaling algorithm) would lead to better practical performance. An implementation that is faster than those of Edmonds' algorithm would also impact the applications in practice.

The \MWM{} problem is fundamental in combinatorial optimization and it appears in many applications including  scheduling, network switching, image analysis, and as a key subroutine in other optimization algorithms. Gabow and Tarjan, for example, discuss its use in Christofides' travelling-salesman approximation and in update problems that require matching duals~\cite{GabowTarjan1991}. Cook and Rohe report minimum-weight perfect-matching computations on geometric instances with as many as five million vertices~\cite{CookRohe1999}. Such uses motivate studying not only asymptotic bounds but also the work that exact weighted-matching programs perform in practice.

In this work, we implement and compare three \MWM{} algorithms, which we typeset in small capitals to distinguish them from the authors and algorithms they are named after. \Edm{} is our implementation of Edmonds' weighted blossom algorithm, organized as in Galil, Micali, and Gabow~\cite{GalilMicaliGabow1986}. \Gab{} runs the same search but maintains its candidate edges with the supporting forest and tree-blossom merging of Gabow's weighted-matching data structures~\cite{Gabow1990,Gabow2018}. \Scal{} implements the Hybrid algorithm of Duan, Pettie, and Su, with its weight scales, inherited blossoms, and bounded searches~\cite{DuanPettieSu2018}. The three implementations share a primal--dual foundation but differ in how they find the next tight edge, how they organize their searches, and how they use the bits of the edge weights.

Implementation studies reveal behaviors of algorithms that asymptotic analysis alone cannot predict~\cite{MehlhornSchaefer2002,Kolmogorov2009}. Our central question is how edge candidate management, weight scaling, and graph structure affect the running time and memory of the three algorithms. We vary graph order, density, weight range, and structure, using Erd\H{o}s--R\'enyi (ER) graphs, bipartite random graphs, stochastic block models, random geometric graphs, rewired rings, and disjoint clique and biclique blocks, which change connectivity and odd-cycle structure. To find graphs on which Gabow's structure pays off, we also built graphs from dense bipartite and odd-cycle structures.

\Edm{} and \Gab{} find a \MWM{}, whereas \Scal{} finds a maximum-weight perfect matching (\MWPM{}). For this reason, we divided our test set into two suites: Suite~A, containing instances that do not admit perfect matchings, and Suite~B, containing instances that do admit perfect matchings. Suite~A is therefore used to compare \Edm{} and \Gab{} on \MWM{} instances, and Suite~B is used to compare all three implementations on graphs whose \MWM{} and \MWPM{} optima are equal.

In this article, we address three research questions.
\begin{enumerate}
  \item \textbf{RQ1: Do Gabow's data structures pay off in practice?} Within an otherwise identical search, does Gabow's supporting-forest candidate structure reduce running time or memory relative to a single lazily-pruned candidate queue, and on which instances?
  \item \textbf{RQ2: When does weight scaling pay off?} How does the Hybrid scaling program compare with \Edm{} and \Gab{}?
  \item \textbf{RQ3: How do instance properties affect performance?} How do graph order, density, weight range, and structure affect running time and memory?
\end{enumerate}

Our contributions are: implementations of Gabow's supporting-forest candidate structure for weighted matching and of the Duan–Pettie–Su Hybrid algorithm, with their departures from the published analyses stated explicitly; a protocol that compares implementations with different objectives only on graphs where the objectives provably agree, and that fixes the tie rule so that \Edm{} and \Gab{} perform the same search; and measurements, with operation counts and phase times, that show where each program spends its time.

The rest of this article is organized as follows. Section~\ref{sec:background} provides preliminaries and definitions. Section~\ref{sec:algorithms} describes the three algorithms. Section~\ref{sec:implementation} describes the implementations. Sections~\ref{sec:validation} and~\ref{sec:methodology} describe the validation and experimental design, Sections~\ref{sec:results} and~\ref{sec:discussion} report and discuss the results. Sections~\ref{sec:related}--\ref{sec:conclusion} review related work, and conclude.

\section{Preliminaries}
\label{sec:background}

\subsection{Matchings}
\label{sec:objectives}

Let $G=(V,E)$ be a simple undirected graph with $n=|V|$ vertices, $m=|E|$ edges, and $w:E \to \mathbb{R}_{\geq 0}$ a nonnegative edge weight function. A matching $M\subseteq E$ is a set of edges no two of which share an endpoint; its cardinality is $|M|$ and its weight is
\begin{equation}
  w(M)=\sum_{e\in M}w(e).
\end{equation}
A maximum weighted matching (\MWM) has the maximum weight among all matchings. A perfect matching covers every vertex, so it has $n/2$ edges and it exists only when $n$ is even. A maximum-weight perfect matching (\MWPM) has the maximum weight among all perfect matchings; the problem is infeasible when $G$ has none. We write
\begin{align}
  \operatorname{OPT}(\MWM)&=\max\{w(M):M\text{ is a matching}\},\\
  \operatorname{OPT}(\MWPM)&=\max\{w(M):M\text{ is a perfect matching}\}.
\end{align}

\begin{figure}[tbp]
  \centering
  \includegraphics{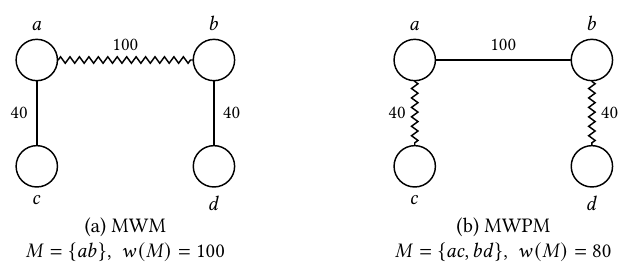}
  \caption{The two objectives can have different optima on the same graph. Wavy edges are matched.}
  \Description{Two copies of a four-vertex graph. On the left, the maximum weighted matching takes one edge of weight 100. On the right, the maximum-weight perfect matching takes two edges whose weights sum to 80.}
  \label{fig:mwm-vs-mwpm}
\end{figure}

Because every perfect matching is a matching, $\operatorname{OPT}(\MWM)\geq\operatorname{OPT}(\MWPM)$, with equality if and only if some perfect matching attains the \MWM{} optimum. The existence of a perfect matching does not guarantee equality: the graph of Figure~\ref{fig:mwm-vs-mwpm} has the perfect matching $\{ac,bd\}$ of weight~80, but its maximum weighted matching is $\{ab\}$, of weight~100. Moreover, when the optima are equal, an \MWM{} solver may still return a non-perfect matching of the same weight.

\subsection{Alternating paths and blossoms}
\label{sec:blossoms}

Relative to a matching $M$ in a graph $G$, a vertex is free if no edge of $M$ is incident to it. An alternating path is a simple path such that its edges alternate between edges in $M$ and edges not in $M$; an augmenting path is an alternating path between two distinct free vertices, and replacing its matched edges by its unmatched ones increases $|M|$ by exactly one. A set of vertices $B$ with $|B|$ odd is a blossom if the vertices in $B$ form a cycle $C$ whose edges alternate between matched and unmatched edges except at one vertex, where two unmatched edges meet, and the number of matched edges on $C$ is $(|B|-1)/2$. The base of the blossom is the unique vertex $b \in B$ that is not incident to any matched edge of $C$; equivalently, $b$ is the vertex at which the blossom attaches to the even-length alternating path which is called a stem, and if a stem does not exist then the base is a free vertex. An alternating search grows a forest whose roots are free vertices or contracted blossoms with a free base. A node of this forest is outer, labeled $\mathrm{S}$, when its distance from the root is even, and inner, labeled $\mathrm{T}$, when it is odd; we also write $S$ and $T$ for the sets of original vertices in outer and inner nodes, and call all other vertices unlabeled. Unlike the two sides of a bipartite graph, two outer nodes may be adjacent: an edge between outer nodes of different trees completes an augmenting path, and an edge between outer nodes of the same tree closes an odd alternating cycle.

\begin{figure}[tbp]
  \centering
  \includegraphics[width=\linewidth]{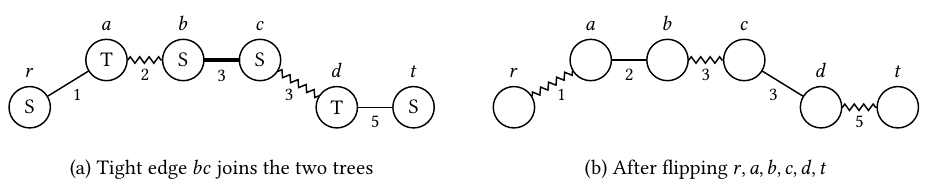}
  \caption{An augmenting path between two alternating trees. $\mathrm{S}$ and $\mathrm{T}$ mark outer and inner vertices; wavy edges are matched.}
  \Description{Two alternating trees rooted at r and t, joined by the edge bc. In panel a, the path r, a, b, c, d, t alternates between unmatched and matched edges. In panel b, flipping the path gives one more matched edge.}
  \label{fig:augmenting-path}
\end{figure}

Figure~\ref{fig:augmenting-path} shows the first case: the edge $bc$ joins outer vertices of the trees rooted at $r$ and $t$, and flipping the path $r,a,b,c,d,t$ adds one edge to the matching.

Contracting an odd cycle closed within one tree into a single outer node, a blossom, lets the search continue in the contracted graph. Blossoms may be nested, but their family is laminar: any two are disjoint or one contains the other~\cite{Gabow2018,DuanPettieSu2018}. The edge set $E_B$ of a blossom $B$ consists of the edges of the odd cycle that formed $B$ together with the edge sets of its subblossoms~\cite{DuanPettieSu2018}. The contracted graph replaces each maximal blossom, one that lies in no other blossom, by one vertex and keeps the edges between different contracted vertices. When an augmenting path of the contracted graph passes through a blossom, it is lifted to $G$ along the unique even-length alternating path in $E_B$ from the vertex where the path enters to the blossom's base.

\begin{figure}[tbp]
  \centering
  \includegraphics[width=\linewidth]{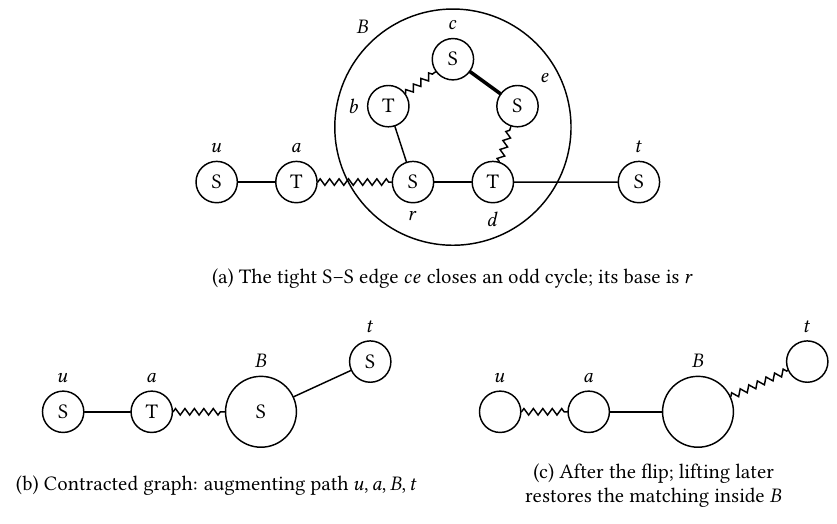}
  \caption{Contraction makes an inner vertex usable. Before contraction, the inner vertex $d$ cannot extend an augmenting path to $t$; in the contracted graph, $u,a,B,t$ is augmenting. Wavy edges are matched.}
  \Description{Three panels. In panel a, the edge ce closes an odd cycle through r, b, c, e and d, which forms the blossom B. In panel b, B is contracted to one outer node, and u, a, B, t is an augmenting path. Panel c shows the graph after the flip.}
  \label{fig:blossom-quotient}
\end{figure}

In Figure~\ref{fig:blossom-quotient}, the edge $ce$ closes a blossom $B$ with base $r$. Before the contraction, the inner vertex $d$ cannot continue an augmenting path to $t$; after it, $B$ is outer and $u,a,B,t$ is augmenting. Lifting routes this path through $B$ along $r,b,c,e,d$, using the edge $ce$ that closed the blossom.

\subsection{Primal--dual formulation}
\label{sec:primal-dual}

Let $\delta(v)$ be the edges incident to $v$, $E(B)$ the edges with both endpoints in $B\subseteq V$, $x(F)=\sum_{e\in F}x(e)$ for a set $F$ of edges, and $\mathcal O$ the family of odd sets $B\subseteq V$ with $|B|\geq3$. With one variable $x(e)$ per edge, the two problems share their objective, the odd-set inequalities and nonnegativity; they differ only at the vertices:
\begin{align}
  \text{maximize}\quad & \sum_{e\in E}w(e)x(e),\\
  x(\delta(v))&\begin{cases}
      \leq 1,&\text{for \MWM,}\\
      =1,&\text{for \MWPM,}
    \end{cases}
    &&\text{for every }v\in V,\\
  x(E(B))&\leq\lfloor|B|/2\rfloor
    &&\text{for every }B\in\mathcal O,\\
  x(e)&\geq0 &&\text{for every }e\in E.
\end{align}
The vertex inequality allows free vertices in \MWM{}, and the equality forces every vertex to be covered in \MWPM{}. Edmonds proved that these systems have integral extreme points, so the constraint $x(e)\in\{0,1\}$ is unnecessary~\cite{EdmondsWeighted1965}. The dual has a variable $y(v)$ for each vertex and a nonnegative variable $z(B)$ for each $B\in\mathcal O$, and it minimizes $\sum_{v\in V}y(v)+\sum_{B\in\mathcal O}\lfloor|B|/2\rfloor\,z(B)$. The \MWM{} dual requires $y(v)\geq0$; the \MWPM{} dual does not restrict the sign of $y(v)$. The dual value of an edge $e=uv$ is
\begin{equation}
  yz(e)=y(u)+y(v)+\sum_{B\in\mathcal O:\,u,v\in B}z(B),
\end{equation}
and in the three algorithms only blossoms have positive $z$; we write $\Omega\subseteq\mathcal O$ for the laminar family of blossoms that an algorithm maintains. Following Galil, Micali and Gabow~\cite{GalilMicaliGabow1986}, Galil~\cite{Galil1986Survey} and Mehlhorn and Sch\"afer~\cite{MehlhornSchaefer2002}, we define the slack of $e$ as
\begin{equation}
  \pi(e)=yz(e)-w(e).
  \label{eq:edge-slack}
\end{equation}
Dual feasibility requires $\pi(e)\geq0$ for every edge, and $e$ is tight when $\pi(e)=0$.

All three algorithms adjust vertex and blossom duals until a tight edge enables a grow step, an augmentation or a blossom contraction, or until the dual of an inner blossom reaches zero and the blossom is expanded. The two objectives remain visible in the control flow. When an outer vertex reaches dual zero, \Edm{} and \Gab{} flip the even alternating path from the root of its tree to that vertex, which leaves the vertex free, and retire the tree. \Scal{} instead makes its matching perfect in its Perfection and Finalization steps, and during intermediate scales it maintains only near domination, $\pi(e)\geq-2$.

\subsection{Algorithmic bounds}

\begin{table}[htbp]
\centering
\caption{Published time bounds of the three algorithms. For Hybrid, the middle bound is per scale.}
\label{tab:published-guarantees}
\small
\begin{tabularx}{\linewidth}{@{}L{0.15\linewidth}L{0.16\linewidth}Y L{0.14\linewidth}L{0.17\linewidth}@{}}
\toprule
\textbf{Algorithm} & \textbf{Objective} & \textbf{Main mechanism} & \textbf{Per search} & \textbf{Total} \\
\midrule
Galil--Micali--Gabow & \MWM & Generalized priority queues, concatenable queues & $O(m\log n)$ & $O(mn\log n)$ \\
Gabow & \MWPM; easily modified for other variants & Supporting forest, NCA with linking, tree-blossom merging, Fibonacci heaps & $O(m+n\log n)$ & $O(n(m+n\log n))$ \\
Duan--Pettie--Su Hybrid & \MWPM; integer weights in $[0,W]$ & Weight scaling, inherited-blossom dismantling, bounded searches & $O(m\sqrt n)$ & $O(m\sqrt n\log(nW))$ \\
\bottomrule
\end{tabularx}
\end{table}

The first row of Table~\ref{tab:published-guarantees} gives the bound of Galil, Micali and Gabow for Edmonds' algorithm, not that of Edmonds' original presentation. Gabow presents his algorithm for maximum-weight perfect matching for definiteness and notes that it is easily modified for the other weighted variants~\cite{Gabow2018}; \Gab{} applies it to \MWM{}. Gabow's algorithm accepts real-valued weights and its time bound is strongly polynomial; for notational simplicity he assumes $m\geq n/2$, which holds once isolated vertices are discarded, and his space bound is $O(m)$. The bound of Duan, Pettie and Su assumes integer weights in $[0,W]$ and a threshold $\tau$ in the range discussed in Section~\ref{sec:scaling-implementation}. Section~\ref{sec:implementation} describes how the measured programs depart from these analyses.

\section{Algorithms}
\label{sec:algorithms}

Edmonds' algorithm (Algorithm~1) and Gabow's algorithm (Algorithm~2) use the state $\mathcal X=(M,\Omega,y,z,\mathcal A)$: the matching, the laminar family of blossoms, the vertex duals, the blossom duals and the alternating forest. Let $B_v$ be the maximal blossom containing $v$, a singleton if $v$ lies in no blossom; let $\operatorname{root}(B_v)$ be the root of the tree that contains $B_v$; and let $\mathcal R$ be the set of active roots. For initialization, $w_{\max}(v)$ is the largest weight of an edge incident to $v$, or zero if there is none.

In the minima below, an edge $uv$ is a candidate only if $B_u\neq B_v$; we leave this condition implicit. The minimum of an empty set is $+\infty$.

\subsection{Edmonds}
\label{sec:edmonds}

\Edm{} maintains a matching, a laminar family of blossoms, dual values and an alternating forest rooted at free maximal blossoms. Tight edges drive the search. An edge from an outer node to an unlabelled node grows the forest, which adds that node to the tree as inner and its mate as outer, or augments if the unlabelled node is free; an edge between two outer nodes augments if they lie in different trees and closes a blossom if they lie in the same tree. An inner blossom whose dual reaches zero is expanded.

When no tight edge enables a step, the duals are adjusted by the smallest of four delta values:
\begin{align}
 \delta_1&=\min_{v\in S}y(v),
 &\delta_2&=\min_{\substack{uv\in E\\u\in S,\ v\text{ unlabelled}}}\pi(uv),\\
 \delta_3&=\tfrac12\min_{\substack{uv\in E\\u,v\in S}}\pi(uv),
 &\delta_4&=\tfrac12\min_{\substack{B\text{ maximal inner}\\|B|\geq3}}z(B).
\end{align}
These delta values enable, respectively, settling a tree, growing or augmenting to an unlabelled node, augmenting or contracting along an edge between outer nodes, and expanding an inner blossom. Settling a tree flips the even alternating path from its root to an outer vertex whose dual is zero, which leaves that vertex free, and retires the tree. The algorithm stops when no active root remains.

\begin{figure}[tbp]
  \centering
  \includegraphics[width=\linewidth]{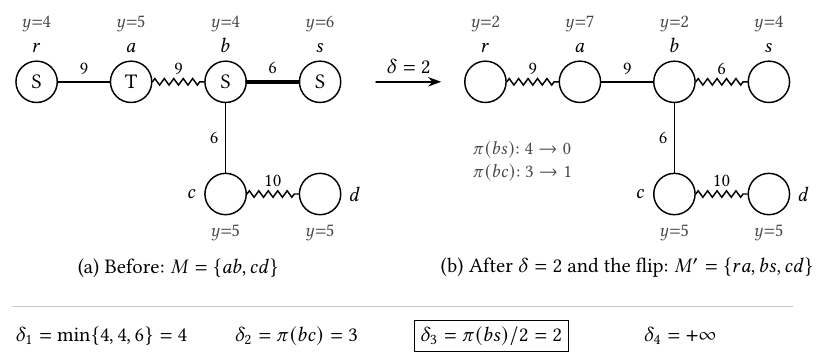}
  \caption{A dual adjustment by $\delta=\delta_3=2$ makes $bs$ tight and enables an augmentation between the trees rooted at the free vertices $r$ and $s$. Plain edge labels are weights; wavy edges are matched.}
  \Description{A graph before and after a dual adjustment of 2. The edge bs becomes tight and the matching changes from ab and cd to ra, bs and cd. A row below the graph lists the four delta values, with the third, equal to 2, chosen.}
  \label{fig:dual-event}
\end{figure}

\begin{figure}[tbp]
  \centering
  \includegraphics[width=\linewidth]{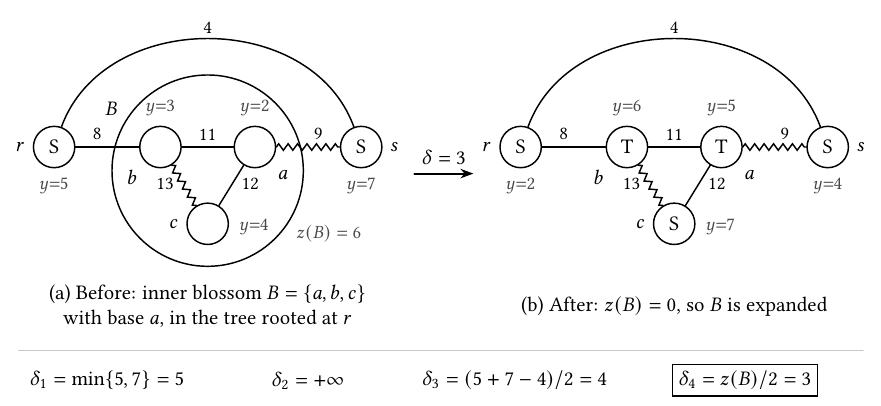}
  \caption{The dual adjustment $\delta=\delta_4=3$ sets $z(B)=0$, and $B$ is expanded. Wavy edges are matched.}
  \Description{An inner blossom B with vertices a, b and c in a tree rooted at r, before and after a dual adjustment of 3. The blossom dual falls from 6 to 0, and B is expanded into its three vertices. A row below the graph lists the four delta values, with the fourth, equal to 3, chosen.}
  \label{fig:blossom-dual-event}
\end{figure}

Figure~\ref{fig:dual-event} shows a step in which $\delta_3=2$ is the smallest delta value: after the dual adjustment, $bs$ is tight, and augmenting along $r,a,b,s$ gives a matching whose weight, 25, equals the dual objective, which certifies its optimality. In Figure~\ref{fig:blossom-dual-event} the smallest delta value is $\delta_4=3$: the dual of the inner blossom reaches zero, so the blossom is expanded and its edges stay tight.

The search preserves four invariants: $M$ is a matching; the blossom family is laminar and every blossom is full (a blossom of $|B|$ vertices is full if $(|B|-1)/2$ of its edges are matched); the duals are feasible; and matched edges and the edges of every $E_B$ are tight. Augmenting along a path $Q$, lifted to the original graph and identified with its set of edges, replaces $M$ by the symmetric difference $M\oplus Q=(M\cup Q)\setminus(M\cap Q)$. Appendix~\ref{app:shared-contract} gives the dual adjustments, the transitions and the optimality certificate.

\Edm{} and \Gab{} keep $\delta_1$, $\delta_2$ and $\delta_4$ in the same priority queues: the duals of the outer vertices for $\delta_1$, the duals of the inner blossoms for $\delta_4$, and, for $\delta_2$, one entry for each unlabelled node, its edge of smallest slack from an outer vertex, which a concatenable queue over the node's vertices finds as in Galil, Micali and Gabow~\cite{GalilMicaliGabow1986}. Only $\delta_3$ is handled differently. \Edm{} and \Gab{} keep the candidates for $\delta_3$ in a candidate structure $\mathcal D$. The query $\operatorname{MinCandidate}(\mathcal D)$ returns an edge $e_3$ with $\delta_3=\pi(e_3)/2$, or $(\bot,+\infty)$ if there is none. The two algorithms differ only in $\mathcal D$ and in the work needed to restore its minimum after the search changes.

Each event has a witness $\xi_i$: a vertex for event~1, an edge for events~2 and~3, and a blossom for event~4. The operation $\operatorname{Dual}$ applies the dual adjustment by $\delta=\min_i\delta_i$, and $\operatorname{Transition}$ performs the step that the chosen event enables. When several delta values are equal, \Edm{} and \Gab{} process the event with the smallest index, update the queues for the vertices and blossoms that the step changed, and read the four minima again before choosing the next event. Ties among the candidates for $\delta_1$, $\delta_2$ and $\delta_4$ are broken by the queues that the two programs share, so only $\delta_3$, whose candidates they keep in different structures, needs an explicit rule: among candidates with equal keys they take the edge $uv$, $u<v$, with the smallest $u$ and, among those, the smallest $v$. They therefore perform the same sequence of search steps (Section~\ref{sec:validation}).

Galil, Micali and Gabow maintain these minima with generalized priority queues and the blossom memberships with concatenable queues~\cite{GalilMicaliGabow1986}. Their queue for $\delta_3$ keeps edges whose endpoints have entered the same blossom and skips them when they reach the top; \Edm{} applies the same lazy discipline to its single candidate queue (Section~\ref{sec:gabow-implementation}). Each edge has at most one current entry in this queue, and giving an edge a new key replaces its earlier entry. An entry is stale when it has been replaced, when its key differs from the current $\pi(e)/2$, or when its edge is no longer a candidate because both endpoints now lie in one blossom or one of them is no longer outer. A stale entry stays in the queue until it reaches the top. It is then discarded, and an edge that is still a candidate is reinserted with its current key unless its current entry already has that key.

\begin{figure}[tbp]
\begin{pseudocode}{Algorithm 1}{Edmonds$(G,w)$}
  \pseudometa{Input}{An undirected graph $G=(V,E)$ with edge weights $w$.}
  \pseudometa{Output}{A maximum weighted matching $M$.}
  \pline{$M,\Omega\leftarrow\varnothing$; $y(v)\leftarrow\tfrac12 w_{\max}(v)$ for every $v\in V$; $z\leftarrow0$.}
  \pline{$\mathcal A\leftarrow$ one search tree per vertex; $\mathcal R\leftarrow V$.}
  \pline{$\mathcal D\leftarrow$ priority queue of the outer--outer candidates, keyed by $\pi(e)/2$.}
  \pline{\textbf{while} $\mathcal R\neq\varnothing$ \textbf{do}}
  \pline[1]{$(e_3,\delta_3)\leftarrow\operatorname{MinCandidate}(\mathcal D)$.}
  \pline[1]{Compute $\delta_1,\delta_2,\delta_4$ and their witnesses.}
  \pline[1]{$\delta\leftarrow\min\{\delta_1,\delta_2,\delta_3,\delta_4\}$; choose $i$ with $\delta_i=\delta$ and its witness $\xi_i$.}
  \pline[1]{$\mathcal X\leftarrow\operatorname{Dual}(\mathcal X,\delta)$.}
  \pline[1]{$\mathcal X\leftarrow\operatorname{Transition}(\mathcal X,i,\xi_i)$.}
  \pline[1]{Insert the new outer--outer candidates into $\mathcal D$; stale entries stay until they reach the top.}
  \pline{\textbf{end while}}
  \pline{\textbf{return} $M$}
\end{pseudocode}
\Description{Pseudocode of the Edmonds program: initialization, then a loop that finds the next candidate, computes the smallest delta value, applies the dual adjustment and the transition, and inserts the new candidates.}
\end{figure}

The initial duals satisfy $y(u)+y(v)\geq w(uv)$ for every edge $uv$ and are nonnegative, so they are feasible. Galil, Micali and Gabow start every vertex at half the largest weight of the graph, so that all free vertices keep equal duals and the algorithm stops when they reach zero~\cite{GalilMicaliGabow1986}. With per-vertex starts, outer vertices of different trees can reach dual zero at different times; each time one does, its tree is settled. Since an augmentation retires at least one tree and a settlement one, there are at most $n$ augmentations and settlements together, against at most $n/2$ augmentations with equal starts, so the per-vertex start at most doubles the number of stages in their analysis. Each transition also updates the active roots $\mathcal R$ and the affected part of the forest. Unlike the stages of Galil, Micali and Gabow, which erase all labels after each augmentation~\cite{GalilMicaliGabow1986}, \Edm{} and \Gab{} keep the trees that an augmentation does not touch, together with their labels. Keeping them saves growing them again after every augmentation.

\subsection{Gabow}
\label{sec:gabow}

\Gab{} uses the same matching, dual adjustments, alternating forest and blossom transitions as \Edm{}; only its candidate structure differs. That structure moves the candidates onto a supporting forest, where each candidate edge is split at the nearest common ancestor of its endpoints, and maintains them with Gabow's tree-blossom merging~\cite{Gabow1990,Gabow2018}.

The alternating forest $\mathcal A$ changes whenever blossoms are contracted or expanded. A separate supporting forest $\mathcal T_{\mathrm{sup}}$ holds every vertex of an outer blossom and, of an inner blossom, the vertices on the path from its entry vertex to its base; its trees hang from a common dummy root $\rho$. For a candidate $e=uv$, let $a$ be the nearest common ancestor (NCA) of $u$ and $v$ in $\mathcal T_{\mathrm{sup}}$. The candidate is replaced by the two ancestor-directed records $(u,a;e)$ and $(v,a;e)$, which keep the original edge and its key $\pi(e)/2$; if $u$ or $v$ is itself the NCA, its record would have no length and is omitted, and a candidate between different trees has $a=\rho$.

Tree-blossom merging partitions $\mathcal T_{\mathrm{sup}}$ into blocks, the sets that Gabow calls the blossoms of the tree-blossom-merging problem. A block $B$ has rank $r(B)=\lfloor\log_2|B|\rfloor$ and keeps its records in lists $\operatorname{packet}(B,k)$, one for each rank $k$, each with a cached minimum; following Gabow, the rank of a record is computed from the logarithm of its path length and from the sizes of the blocks at its two ends~\cite{Gabow2018}. When blocks are merged, only the records whose rank is at most that of the new block are examined again; the other packets are joined whole. Newly classified records wait in a list $\operatorname{loose}(B)$ until a merge places them in a packet. Packet and loose-list minima feed the minimum of each block, and block minima feed the global minimum. Records are validated at the original endpoints of their edges.

\begin{figure}[tbp]
  \centering
  \includegraphics[width=\linewidth]{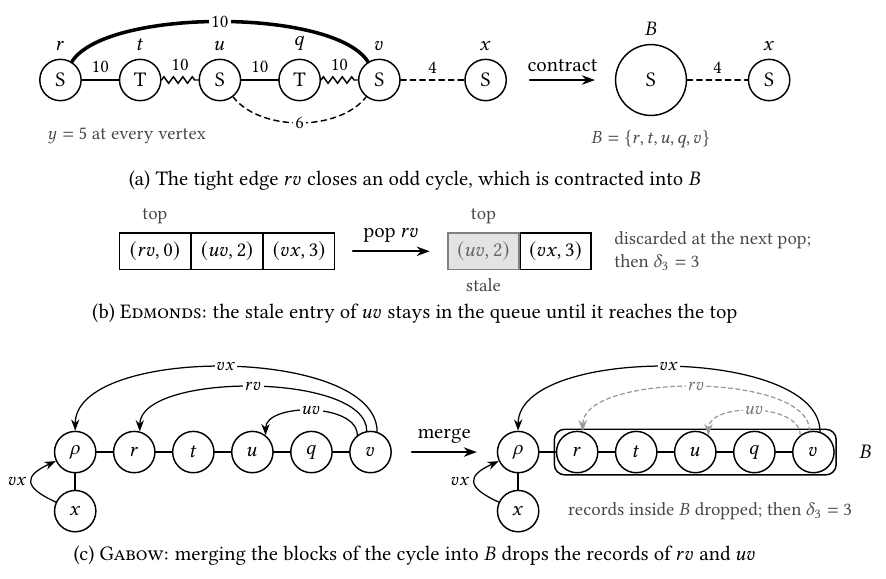}
  \caption{Candidate maintenance after a contraction. Plain edge labels are weights, dashed edges are candidates, wavy edges are matched, and curved arrows are records.}
  \Description{Three panels. In panel a, the tight edge rv closes an odd cycle that is contracted into the blossom B. In panel b, the queue of Edmonds holds entries for rv, uv and vx; the entry of uv becomes stale and stays until it reaches the top. In panel c, the supporting tree of Gabow has curved arrows for records; merging the blocks of the cycle into B drops the records of rv and uv.}
  \label{fig:gabow-candidates}
\end{figure}

Figure~\ref{fig:gabow-candidates} shows the two structures on one contraction, with each record $(u,a;e)$ drawn as a curved arrow from $u$ to $a$. When the tight edge $rv$ closes an odd cycle, the candidate $uv$ becomes internal to the new blossom: \Edm{} discards its stale queue entry only when the entry reaches the top, whereas \Gab{} drops its record while merging the blocks of the cycle.

\begin{figure}[tbp]
\begin{pseudocode}{Algorithm 2}{Gabow$(G,w)$}
  \pseudometa{Input}{An undirected graph $G=(V,E)$ with edge weights $w$.}
  \pseudometa{Output}{A maximum weighted matching $M$.}
  \pline{$M,\Omega\leftarrow\varnothing$; $y(v)\leftarrow\tfrac12 w_{\max}(v)$ for every $v\in V$; $z\leftarrow0$.}
  \pline{$\mathcal A\leftarrow$ one search tree per vertex; $\mathcal R\leftarrow V$.}
  \pline{$\mathcal D\leftarrow$ supporting forest, blocks and records of the outer--outer candidates.}
  \pline{\textbf{while} $\mathcal R\neq\varnothing$ \textbf{do}}
  \pline[1]{$(e_3,\delta_3)\leftarrow\operatorname{MinCandidate}(\mathcal D)$.}
  \pline[1]{Compute $\delta_1,\delta_2,\delta_4$ and their witnesses.}
  \pline[1]{$\delta\leftarrow\min\{\delta_1,\delta_2,\delta_3,\delta_4\}$; choose $i$ with $\delta_i=\delta$ and its witness $\xi_i$.}
  \pline[1]{$\mathcal X\leftarrow\operatorname{Dual}(\mathcal X,\delta)$.}
  \pline[1]{$\mathcal X\leftarrow\operatorname{Transition}(\mathcal X,i,\xi_i)$.}
  \pline[1]{Update the supporting positions of vertices that were added, relabelled or retired.}
  \pline[1]{\textbf{if} a blossom was contracted \textbf{then}}
  \pline[2]{\textbf{for each} pair of blocks $X,Y$ merged into $Z$ along the cycle path, bottom-up, \textbf{do}}
  \pline[3]{$\operatorname{packet}(Z,k)\leftarrow\operatorname{packet}(X,k)\cup\operatorname{packet}(Y,k)$ for every $k>r(Z)$.}
  \pline[3]{Reclassify the other records of $X$ and $Y$; drop those internal to $Z$ and all but the smallest between $Z$ and each other block.}
  \pline[2]{\textbf{end for}}
  \pline[1]{\textbf{else if} a blossom $B$ was expanded \textbf{then}}
  \pline[2]{For each child of $B$ that becomes outer, add its vertices that are missing from $\mathcal T_{\mathrm{sup}}$ below its base, and merge its blocks into one.}
  \pline[1]{\textbf{end if}}
  \pline[1]{\textbf{for each} affected edge $e=uv$ \textbf{do}}
  \pline[2]{Remove the old records of $e$.}
  \pline[2]{\textbf{if} $u,v\in S$ and $B_u\neq B_v$ \textbf{then} insert $(u,a;e)$ and $(v,a;e)$, where $a=\operatorname{nca}(u,v)$, with key $\pi(e)/2$.}
  \pline[1]{\textbf{end for}}
  \pline[1]{Refresh the minima of the changed packets, loose lists and blocks.}
  \pline{\textbf{end while}}
  \pline{\textbf{return} $M$}
\end{pseudocode}
\Description{Pseudocode of the Gabow program, with the steps of Algorithm 1 except in two places. Line 3 builds the supporting forest, blocks and records. Lines 10 to 23 update the supporting positions, the blocks after a contraction or an expansion, and the records of the affected edges.}
\end{figure}

Algorithm~2 differs from Algorithm~1 in two places only: line~3 builds the supporting forest, blocks and records instead of a queue, and lines~10--23 replace line~10 of Algorithm~1. An edge is affected when it becomes a candidate, or when a change of tree, vertex position or block invalidates or reprices its records. New records enter the packet or loose list given by their owning block and rank. The logical key of a record is $\pi(e)/2$; the implementation stores it with lazy offsets.

With his analysed structures, Gabow's search takes $O(m+n\log n)$ time, and at most $n/2$ searches give $O(n(m+n\log n))$ in total~\cite{Gabow2018}. Section~\ref{sec:gabow-implementation} lists where \Gab{} departs from these structures.

\subsection{Scaling}
\label{sec:scaling}

\Scal{} solves \MWPM{} on general graphs with nonnegative integer weights. Let $\widehat w$ be the input weight and $\bar w(e)=(n/2+1)\widehat w(e)$ the extended weight. Under $\bar w$, the weight of every matching is a multiple of $n/2+1$, so the additive error that the relaxed scales allow cannot change the optimum (Appendix~\ref{app:scaling-invariants})~\cite{DuanPettieSu2018}.

\begin{figure}[tbp]
  \centering
  \includegraphics{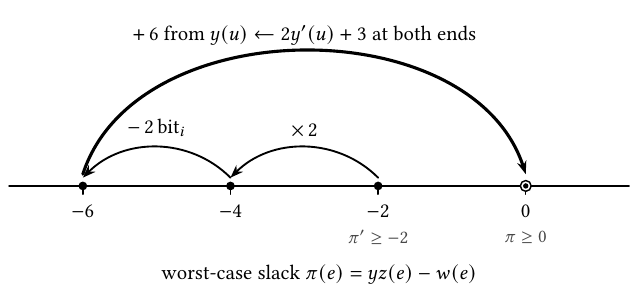}
  \caption{Worst-case slack at a scale transition. Near domination ($\pi'\geq-2$) becomes $-4$ after doubling and $-6$ after the new weight bit; adding 3 at both endpoints restores $\pi\geq0$.}
  \Description{Worst-case edge slack at a scale transition: minus 2 before the transition becomes minus 4 after doubling and minus 6 after the new weight bit, and adding 3 at both endpoints brings it back to 0.}
  \label{fig:scale-slack}
\end{figure}

At the start of each scale, \Scal{} reveals the next bit of $\bar w$ and updates the weights and duals $(w',y',z')$ of the previous scale by
\begin{equation}
  w(e)\leftarrow2\bigl(w'(e)+\operatorname{bit}_i(\bar w(e))\bigr),\qquad
  y(u)\leftarrow2y'(u)+3,\qquad
  z(B)\leftarrow2z'(B).
\end{equation}
Doubling keeps the old dual certificate valid for the doubled weights. The previous scale guarantees only near domination, $\pi'(e)\geq-2$; doubling and the new bit lower the worst-case slack to $-6$, and adding~3 to both endpoint duals restores $\pi(e)\geq0$ (Figure~\ref{fig:scale-slack}).

A threshold $\tau$ divides inherited blossoms into large ones, with $|B|\geq\tau$, and small ones. \Scal{} uses $\tau=s\lceil\log_2(\max\{2,n\})\rceil$ with $s=\lceil\sqrt n\rceil$; both values also serve as the dual budgets of free-vertex reduction, and Section~\ref{sec:scaling-implementation} discusses the choice. Liquidating a blossom $B$ adds $z(B)/2$ to $y(u)$ for every $u\in B$, sets $z(B)=0$ and removes $B$ from its family; this leaves $yz(e)$ unchanged for every edge inside $B$ and increases it by $z(B)/2$ for every edge with exactly one endpoint in $B$.

The edge set $E_B$ of a blossom (Section~\ref{sec:blossoms}) supports base changes and path lifting; by contrast, $E(B)$ contains every input edge with both endpoints in $B$. Once the inherited blossoms are gone, \Scal{} maintains relaxed complementary slackness~\cite{DuanPettieSu2018}: every blossom is full, every maximal blossom of the current family has positive dual, the edges of $M$ and of every $E_B$ satisfy $yz(e)\leq w(e)$, and every edge satisfies $\pi(e)\geq-2$. While inherited and current blossoms coexist, separate coexistence conditions govern them. Appendix~\ref{app:scaling-invariants} states these conditions and the error bound.

Each scale keeps the duals and blossoms of the previous scale as the inherited family $\Omega'$, empties the matching, and starts a new current family $\Omega$~\cite{DuanPettieSu2018}. No matched edge crosses an inherited blossom while that blossom is processed, and no current blossom contains an inherited one.

\Scal{} first liquidates every large inherited blossom, including large blossoms nested in other large ones. It then decomposes each maximal small inherited blossom with positive dual into major paths, which it processes in postorder with \textsc{DismantlePath}. \textsc{DismantlePath} searches the shells between consecutive undissolved inherited boundaries, translates dual adjustments across those boundaries, and expands current blossoms whose dual reaches zero before an inherited boundary changes. A shell search grows alternating forests, augments within the shell, contracts the blossoms it reaches, and adjusts duals until it augments, reaches a shell searched earlier, or the outermost undissolved boundary dissolves. Appendix~\ref{app:scaling-shells} defines the shells and the translations.

Free-vertex reduction (FVR) follows. FVR first runs Criterion-2 phases, each of which augments along a maximal set of vertex-disjoint augmenting paths, until $s$ dual units have elapsed or the matching is perfect; an interval without events may be processed in one step, with its whole length counted. It then runs Criterion-3 searches until the total dual adjustment of FVR reaches $\tau$ or no real vertex is free. Both budgets count units of dual adjustment, not augmentations: one unit lowers the dual of every outer vertex by one and raises that of every inner vertex by one, and any number of augmentations may occur between two units. Perfection deletes the free dummy vertices left by earlier scales and gives each remaining free real vertex $u$ a new dummy mate $\hat u$, joined to $u$ by a matched edge of weight zero. After the last scale, Finalization deletes all dummies and completes a perfect matching of the real graph.

The three search criteria differ in their eligible edges. Criterion~1 admits only tight edges. Criterion~2 admits the edges of current blossoms, unmatched edges with slack $-2$ and matched edges with slack $0$. Criterion~3 admits every edge with slack $0$ or $-2$, matched or not. Augmenting along a path changes which of its edges are matched but not their slacks, so under Criterion~2 the path's edges between blossoms become ineligible, and once a phase has augmented along a maximal set of vertex-disjoint paths, no eligible augmenting path remains (Appendix~\ref{app:scaling-shells}). Each Criterion-3 search augments along one path. Duan, Pettie and Su run every search with one procedure, $\operatorname{PQSearch}$, an implementation of Edmonds' search with a priority queue; the phase that calls it chooses the roots, the criterion and when to stop.

Algorithm~3 summarizes one scale. In Algorithm~3, $\operatorname{bitlength}(0)=1$, so an all-zero weighting still has one scale. The state $\mathcal X$ comprises $M$, $y$, $z$, $w$, $\Omega$ and any undissolved inherited state, and $F(\mathcal X)$ is the set of free real vertices. The call $(\mathcal X,\eta)\leftarrow\operatorname{Search}_c(\mathcal X,F,b)$ runs Criterion-$c$ searches with a budget of $b$ dual units and returns the number $\eta\leq b$ of units it used. A family $\mathcal Q$ of vertex-disjoint augmenting paths, lifted to the original graph, updates the matching by $M\leftarrow M\oplus\bigcup_{Q\in\mathcal Q}Q$. The budget $\Delta$ accumulates across the switch from Criterion~2 to Criterion~3 and is reset only at the next scale.

\begin{figure}[tbp]
  \centering
  \includegraphics[width=\linewidth]{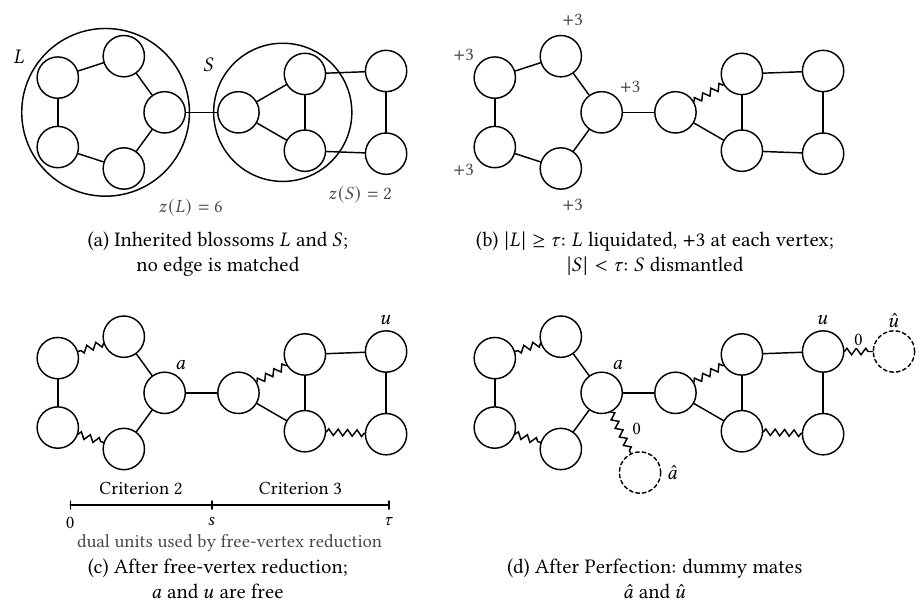}
  \caption{One scale of \Scal{} on a ten-vertex example with $\tau=5$. Wavy edges are matched; the budget axis is schematic.}
  \Description{Four panels of a ten-vertex graph over one scale. Panel a shows the inherited blossoms L and S with no edge matched. In panel b, L is liquidated, adding 3 at each of its vertices, and S is dismantled. Panel c shows the matching after free-vertex reduction, with a budget axis from 0 through s to tau; the vertices a and u are free. In panel d, after Perfection, a and u have dummy mates.}
  \label{fig:scaling-scale}
\end{figure}

Figure~\ref{fig:scaling-scale} shows the steps of one scale on a ten-vertex example. It uses $\tau=5$ in place of the value 16 that the formula gives for $n=10$, so that one inherited blossom is large and one is small.

Each phase preserves an invariant. Dismantling preserves the coexistence conditions until no inherited blossom remains; FVR preserves relaxed complementary slackness; Perfection makes the matching perfect on the augmented graph and preserves the same conditions (Appendix~\ref{app:perfection}); and Finalization returns a perfect matching of the real graph that satisfies them, which implies optimality (Appendix~\ref{app:scaling-invariants}).

\begin{figure}[tbp]
\begin{pseudocode}{Algorithm 3}{Scaling$(G,\widehat w)$}
  \pseudometa{Input}{An even-order graph with nonnegative integer weights and a perfect matching.}
  \pseudometa{Output}{A maximum-weight perfect matching of the real graph.}
  \pline{$s\leftarrow\lceil\sqrt n\rceil$; $\tau\leftarrow s\lceil\log_2(\max\{2,n\})\rceil$; $\bar w\leftarrow(n/2+1)\widehat w$.}
  \pline{$\ell\leftarrow\operatorname{bitlength}(\max_e\bar w(e))$; $M,\Omega\leftarrow\varnothing$; $w,y,z\leftarrow0$.}
  \pline{\textbf{for} $i=\ell-1,\ldots,0$ \textbf{do}}
  \pline[1]{$(w',y',z')\leftarrow(w,y,z)$; $\Omega'\leftarrow\Omega$; $M,\Omega\leftarrow\varnothing$.}
  \pline[1]{$w(e)\leftarrow2(w'(e)+\operatorname{bit}_i(\bar w(e)))$; $y(u)\leftarrow2y'(u)+3$; $z(B)\leftarrow2z'(B)$.}
  \pline[1]{Liquidate every inherited $B$ with $|B|\geq\tau$, including nested large blossoms.}
  \pline[1]{Dismantle the remaining small inherited blossoms by $\textsc{DismantlePath}(R)$ for the major-path roots $R$, in postorder.}
  \pline[1]{$\Delta\leftarrow0$.}
  \pline[1]{\textbf{for} $(c,b)=(2,s),(3,\tau)$, in this order, \textbf{do}}
  \pline[2]{\textbf{while} $F(\mathcal X)\neq\varnothing$ and $\Delta<b$ \textbf{do}}
  \pline[3]{$(\mathcal X,\eta)\leftarrow\operatorname{Search}_c(\mathcal X,F(\mathcal X),b-\Delta)$.}
  \pline[3]{$\Delta\leftarrow\Delta+\eta$.}
  \pline[2]{\textbf{end while}}
  \pline[1]{\textbf{end for}}
  \pline[1]{$\mathcal X\leftarrow\operatorname{Perfection}(\mathcal X)$.}
  \pline{\textbf{end for}}
  \pline{$\mathcal X\leftarrow\operatorname{Finalization}(\mathcal X)$.}
  \pline{\textbf{return} $M\cap E$.}
\end{pseudocode}
\Description{Pseudocode of the Scaling program: for each scale, update the weights and duals, liquidate the large inherited blossoms, dismantle the small ones, run the Criterion-2 and Criterion-3 searches of free-vertex reduction, and apply Perfection. Finalization follows the last scale.}
\end{figure}

\begin{figure}[tbp]
  \centering
  \includegraphics[width=\linewidth]{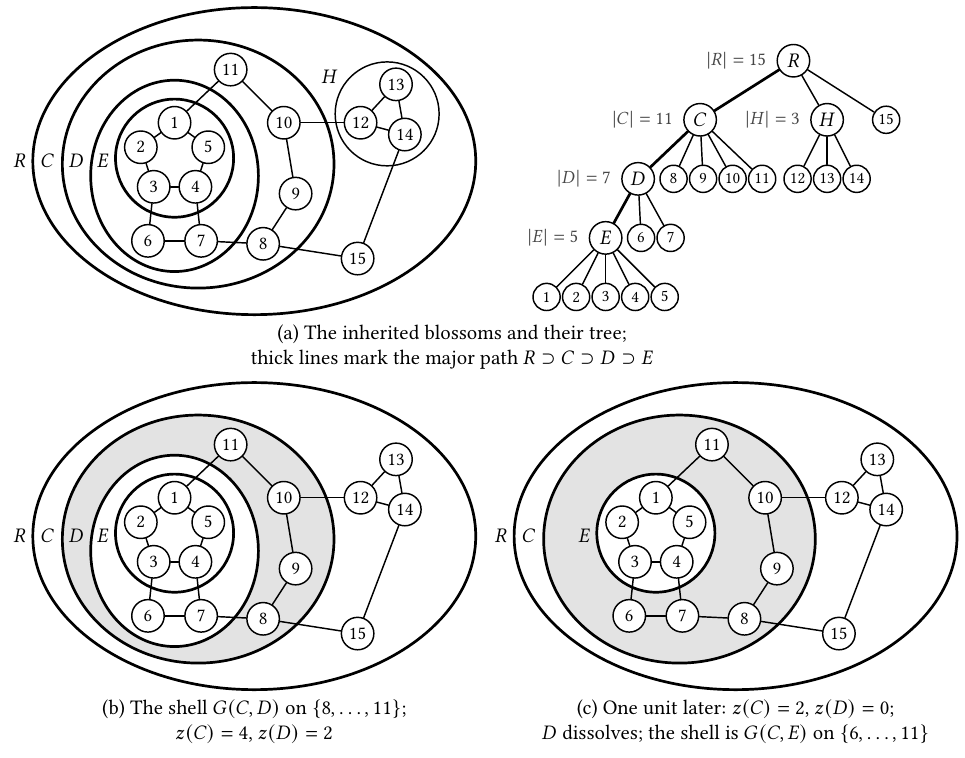}
  \caption{A major path and its shells on a 15-vertex example. Shading marks the shell.}
  \Description{Panel a shows a 15-vertex graph with nested inherited blossoms R, C, D, E and H, and their tree of children, with thick lines along the major path R, C, D, E. Panel b shades the shell between C and D on the vertices 8 to 11. In panel c, one unit of dual adjustment later, D has dissolved and the shell between C and E covers the vertices 6 to 11.}
  \label{fig:major-path-shell}
\end{figure}

The children of a blossom are the vertices and blossoms of the odd cycle that formed it. A child of an inherited blossom is major if it contains more than half of its parent's vertices; starting from a blossom that is not itself a major child and repeatedly moving to the major child gives a major path. Figure~\ref{fig:major-path-shell}a shows a 15-vertex example: each inherited blossom is drawn around its cycle, and the thick edges of the tree of children form the major path $R\supset C\supset D\supset E$. The blossom $H$ starts a path of its own and is dismantled first, because major paths are processed in postorder. Consecutive undissolved boundaries $D\subset C$ on a major path delimit a shell $G(C,D)$, the subgraph induced by $C\setminus D$. Each unit of dual adjustment in a shell lowers the dual of each of its boundaries by two, and a boundary whose dual reaches zero dissolves; the shell then extends to the next undissolved boundary inside it. In Figure~\ref{fig:major-path-shell}b and~c, one unit dissolves $D$, and the shell $G(C,D)$, on the vertices $\{8,\ldots,11\}$, becomes $G(C,E)$, on $\{6,\ldots,11\}$. Appendix~\ref{app:scaling-control} gives the complete control flow.

Duan, Pettie and Su obtain $O(m\sqrt n\log(nW))$ time with their data structures and phase bounds~\cite{DuanPettieSu2018}; Section~\ref{sec:scaling-implementation} describes the structures that \Scal{} uses.

\section{Implementation}
\label{sec:implementation}

The bounds of Table~\ref{tab:published-guarantees} hold for the published algorithms with their analysed data structures and parameters. The measured programs depart from these in the ways described below, and we do not claim the bounds for them. For each departure, here and in Section~\ref{sec:edmonds}, we give its reason and, where we measured it, its cost. The source code of the three implementations is available at \url{https://github.com/AzizFaraj/maximum-weight-matching}, together with the scripts that run and analyse the experiments and the per-trial records.

\subsection{Common representation and candidate structures}
\label{sec:gabow-implementation}
Each implemented algorithm returns a mate vector, the matching cardinality and the matching weight. Recorded running times do not include reading the input, building the graph, or checking the output.

\Edm{} and \Gab{} use the same mate storage, dual offsets, blossom bookkeeping, path lifting, growth, expansion and result extraction. Both release the membership arrays of nested blossoms once an enclosing blossom absorbs them, and both materialize blossom contents iteratively rather than recursively. When a blossom is contracted, the programs build an explicit list of its vertices and a map, ordered by vertex, from each vertex to the child that contains it, so the time of a contraction grows with the size of the new blossom; when one child holds at least half of the vertices, its ordered map is reused and only the other vertices are sorted. Galil, Micali and Gabow instead join the concatenable queues of the children without listing their vertices (Section~\ref{sec:edmonds}). The lists are a simplification common to both programs, which list the same vertices (Table~\ref{tab:counts}). Contraction took at most 2.6\% of the time of \Edm{} on 40 of the 45 profiled graphs, but 19--59\% on the other five, whose contractions list more than a billion vertex entries (Section~\ref{sec:profiles}). \Edm{} keeps the outer--outer candidates in one global priority queue and checks an entry's endpoint labels, trees, blossom membership and slack only when it reaches the top. \Gab{} replaces this queue by the supporting forest, the NCA reduction and tree-blossom merging. The rest of the search and the tie rule of Section~\ref{sec:edmonds} are common, so the two implementations perform the same search and differ in candidate management: the candidate structure, the work of keeping it current and the shared operations that this work calls.

\Gab{} departs from Gabow's analysed structures in three ways. It answers NCA queries by binary lifting, which needs $O(\log n)$ time and space for each inserted vertex and $O(\log n)$ time per query; for the $O(n)$ insertions and $O(m)$ queries of a search this costs $O((n+m)\log n)$. Gabow's structure for a tree that grows by adding leaves answers each query in $O(1)$ time and needs $O(n+m)$ for a search~\cite{Gabow2017,Gabow2018}, which keeps the search within his bound of $O(m+n\log n)$. With binary lifting, the bound of a search becomes $O((m+n)\log n)$, the order of a search of Galil, Micali and Gabow, so this departure gives up the asymptotic advantage of Gabow's structures. In practice this cost is small: on the two graphs where we timed them, including the profiled graph with the most NCA queries, NCA queries and their maintenance took at most 4.1\% of the implemented algorithm time (Section~\ref{sec:profiles}). Its packets and loose lists are pairing heaps; the global heap of block minima is a Fibonacci heap, as in Gabow's design. A pairing heap joins two packets by one link and keeps its minimum at the root, so Gabow's operations take the same time as on his lists with cached minima; it can also remove a single record, which the continuous search needs when it retires records. Its search is continuous: rather than rebuilding the supporting forest for each search, it retires and republishes records when the persistent search changes the supporting position of a vertex, so Gabow's per-search accounting does not apply directly. \Gab{} keeps the search continuous because it shares the search of \Edm{} (Section~\ref{sec:edmonds}), so that the two programs perform the same search steps; the cost of retiring records is included in the measured cost of its candidate management (Section~\ref{sec:profiles}). RQ1 therefore tests the practical value of Gabow's candidate structure as implemented, not his asymptotic improvement.

\Edm{} and \Gab{} store weights and duals as double-precision numbers, which represent the integers and half-integers that arise here exactly within their range; \Scal{} uses checked 64-bit integer arithmetic. The duals of \Edm{} and \Gab{} are stored lazily: each labelled blossom records the running total of the dual adjustments when it was labelled, so a dual adjustment changes one global value instead of every vertex dual. Their dual checks use tolerances of $10^{-5}$ for feasibility and $10^{-4}$ for tightness and for the agreement of the primal and dual objectives.

\subsection{Scaling}
\label{sec:scaling-implementation}

\Scal{} implements Hybrid with a persistent per-edge event and grow frontier in place of the split-findmin grow driver of Duan, Pettie and Su~\cite{DuanPettieSu2018}. It keeps the inherited and current blossom families, large-first liquidation, major-path dismantling, two-stage FVR, Perfection and Finalization. Its data structures include path-compressed union--find, AVL-based split-findmin for the remaining inner-boundary operations, a binary heap for searches without a budget bound, and explicitly stored shell translations. We chose these structures for simplicity. The split-findmin structure performed no split and no key decrease on any of the 50 Suite~B graphs, and its initialization took at most 2.2\% of the time on the 17 graphs on which \Scal{} was profiled; at most 3.2\% of the queue insertions went to the binary heap. Instead of splitting lists, as the grow driver does, the frontier rebuilds the event state of a search after a blossom is expanded or dissolved; this took at most 0.2\% of the time on 16 of the profiled graphs, but 11\% on the toroidal graph (Section~\ref{sec:profiles}). For every vertex it stores the list of inherited blossoms that contain it, from which it computes the inherited blossom duals of an edge; these lists need space proportional to the total nesting depth of the vertices. They are built at the start of each scale and keep the large blossoms after liquidation, with dual zero. This representation is a simplification: on the 47 Suite~B graphs whose inherited blossoms nest at most 359 levels deep, \Scal{} needed less memory than \Edm{}, but on the toroidal graphs, nested up to 60,500 levels deep, the lists take nearly all of its memory (Section~\ref{sec:profiles}). Repeated maximal-augmentation phases compute the maximum-cardinality matchings that shell searches require. The analysis of Duan, Pettie and Su bounds the time of these computations, which we do not claim for ours; on the 44 Suite~B graphs with shell matchings, a computation took at most 1.45 phases on average, counting the final phase that finds no augmenting path. Processing an interval without events in one step (Section~\ref{sec:scaling}) gives the same state as processing its units one at a time.

Perfection follows Duan, Pettie and Su with one rule made explicit. Duan, Pettie and Su omit the reweighting step of their Liquidationist algorithm from Hybrid but perform Perfection as in Liquidationist, which gives each dummy the dual $\tau$; that value makes the dummy edge tight only when $y(u)=-\tau$, which reweighting guarantees in Liquidationist~\cite{DuanPettieSu2018}. \Scal{} sets $y(\hat u)=-y(u)$, which makes the dummy edge tight for every free vertex~$u$. Appendix~\ref{app:perfection} shows that this insertion preserves relaxed complementary slackness when the incoming state satisfies it.

\Scal{} uses $s=\lceil\sqrt n\rceil$, the dual budget of the Criterion-2 phases, and $\tau=s\lceil\log_2(\max\{2,n\})\rceil$, the total dual budget of FVR and the threshold for large blossoms (Section~\ref{sec:scaling}). A larger $\tau$ allows more dual progress during FVR and changes which inherited blossoms count as large. Duan, Pettie and Su derive their bound for $\sqrt n\log n\leq\tau\leq n^{2/3}$~\cite{DuanPettieSu2018}. This interval is empty unless $\log n\leq n^{1/6}$, which with base-2 logarithms requires $n$ above about $10^9$ (with natural logarithms, about $2.4\times10^7$), so no choice of $\tau$ lies in it at our orders. For example, $(n,\tau,n^{2/3})$ is approximately $(250000,9000,3969)$, $(10^6,20000,10000)$ and $(2\times10^6,29715,15874)$. We use this $\tau$ without claiming that it is optimal or that their bound holds for it. We did not vary it, so its effect on the results, in particular on the geometric graphs of Section~\ref{sec:profiles}, remains open. The number of scales is the bit length of the largest extended weight $(\lfloor n/2\rfloor+1)\,W$; for $n=500{,}000$ it is 28 with $W=10^3$ and 38 with $W=10^6$.

\section{Correctness Validation}
\label{sec:validation}

Optimality is checked independently with LEMON~\cite{LEMON2011}, which serves only as a correctness oracle: its maximum weighted matching solver gives the reference \MWM{} weight, and its maximum-weight perfect matching solver gives the reference \MWPM{} weight. LEMON's running times are not part of our comparisons.

In Suite~B, a graph is admitted when LEMON's two solutions are valid and have equal weight; the measured programs are checked afterwards, and their results do not affect admission.

In Suite~A, LEMON's \MWM{} solver was run on all 49 graphs. Every output of the measured programs attains LEMON's optimum weight in both suites. On 12 Suite~A graphs LEMON's matching has a different cardinality at the same weight, which \MWM{} permits.

Untimed instrumented runs, one pair per graph, check that \Edm{} and \Gab{} perform the same search. On all 99 graphs of the two suites, the two programs recorded the same number of steps of each kind and the same step sequences, and returned identical matchings.

The development builds are tested on small instances that are checked exhaustively, with primal--dual certificates, by comparing event-driven and scanning dual adjustments, and on targeted blossom, shell and arithmetic cases. The Perfection argument of Appendix~\ref{app:perfection} assumes a valid incoming state, which we do not verify at every scale boundary of \Scal{}.

We used the generative AI coding assistants Claude Code (Anthropic) and Codex (OpenAI) to help write, test and debug the three programs and the scripts that run the experiments, analyse their records and draw the figures; to help search for the constructions of the structured graphs (Section~\ref{sec:suite-a}); and to help draft and edit the text. We reviewed and checked all of their output and take full responsibility for the content of this article.

\section{Experimental Design}
\label{sec:methodology}
In this section, we describe our test sets, experimental setup, and the comparison methodology.
\subsection{Objectives and test set} %

\begin{table}[htbp]
\centering
\caption{The two experimental suites.}
\label{tab:suites}
\small
\begin{tabularx}{\linewidth}{@{}L{0.08\linewidth}L{0.27\linewidth}Y L{0.20\linewidth}@{}}
\toprule
\textbf{Suite} & \textbf{Programs} & \textbf{Admission} & \textbf{Objective}\\
\midrule
A & \Edm, \Gab & Selected archive and generated graphs; structured graphs built for the comparison & \MWM\\
B & \Edm, \Gab, \Scal & LEMON's \MWM{} and \MWPM{} solutions are valid and have equal weight & \MWM{} $=$ \MWPM\\
\bottomrule
\end{tabularx}
\end{table}

Suite~A compares \Edm{} and \Gab{} on \MWM{}, so vertices may remain free. Suite~B compares all three implementations on even-order graphs: a graph enters Suite~B if and only if LEMON's \MWM{} and \MWPM{} solutions are both valid and $\operatorname{OPT}(\MWM)=\operatorname{OPT}(\MWPM)$.

\subsection{Suite A}
\label{sec:suite-a}

Suite~A contains 49 graphs. Thirteen come from the SuiteSparse Matrix Collection~\cite{DavisHu2011}: four rectangular matrices, which enter as bipartite graphs through their biadjacency form, and nine symmetric matrices: six from the 10th DIMACS Implementation Challenge~\cite{Bader2013}, \texttt{com-LiveJournal} from SNAP~\cite{LeskovecKrevl2014}, and \texttt{kmer\_V2a} and \texttt{nlpkkt200}. Twelve are generated: eight random bipartite graphs (\texttt{erb}), stored as square adjacency matrices, three Barab\'asi--Albert graphs~\cite{BarabasiAlbert1999} (\texttt{ab}) and one ER graph (\texttt{er1}). The other 24 are structured graphs, described below. All graphs use integer weights drawn uniformly at random from the interval $[1,1000]$ and stored in their corresponding files. Table~\ref{tab:suite-a-results} lists the graphs and their dimensions.

\emph{Structured graphs.} The 24 structured graphs have about one million vertices each: two topology seeds for each of twelve constructions. We designed them after measuring the other graphs, to look for graphs on which Gabow's candidate structure pays off: their constructions are those on which \Gab{} was faster in exploratory runs on smaller graphs. Constructions on which \Edm{} was faster in those runs, such as banded and hub-dominated bipartite graphs, were not included, and we report every graph built from the chosen constructions. The five odd-cycle constructions were chosen from runs in which edge weights were sums of vertex weights; in exploratory runs with per-edge weights, as in the structured graphs, \Edm{} was faster on all five, on graphs with 16,320--33,024 vertices.

Most constructions repeat a small dense graph, which we call a block; Table~\ref{tab:structured} describes each construction. Seven constructions are bipartite: a small left side, about 3\% of the vertices, is joined densely to a large right side, so a maximum matching covers only about 6\% of the vertices. The other five consist of blocks with an odd number of vertices, which force blossoms; a maximum matching leaves about one vertex of each block free, and fewer in the clique chain. The vertices of a block are numbered consecutively, and those of a cycle in order around it, so the neighbours of a vertex have nearby numbers.

\begin{table}[htbp]\centering\small
\caption{The twelve constructions of the structured graphs.}
\label{tab:structured}
\begin{tabularx}{\linewidth}{@{}l Y@{}}
\toprule
\textbf{Construction} & \textbf{Description}\\\midrule
\multicolumn{2}{@{}l}{\emph{Bipartite}}\\
Biclique blocks & 237 separate blocks; each of 124--132 left vertices is joined to all 4,096 right vertices of its block\\
Connected blocks & biclique blocks in which two edges inside each pair of consecutive blocks are replaced by two edges between them, which connects the graph without changing any degree\\
Biclique ring & blocks in a ring; half of the left vertices of each block are joined to the first half of its right vertices and to the second half of those of the next block\\
Ferrers blocks & in each block, a left vertex is joined to the first $k$ right vertices, with $k$ growing from 2,048 to 4,096 over the left vertices\\
Convex windows & in each block, a left vertex is joined to a window of 2,048 consecutive right vertices; the window moves across the block from one left vertex to the next\\
Circular intervals & 30,303 left vertices, each joined to 4,096 consecutive vertices of a cycle of 969,696 right vertices\\
Cube neighbourhoods & 4,096 groups of 6--10 left and 232--240 right vertices, one for each corner of a 12-dimensional cube; a left vertex is joined to the right vertices of its group and of the 12 groups at adjacent corners\\
\midrule
\multicolumn{2}{@{}l}{\emph{Odd-cycle blocks}}\\
Odd cliques & separate complete graphs on 125--133 vertices\\
Clique chain & the same cliques, each joined to the next by one edge\\
Tripartite & separate complete tripartite graphs with parts of 41, 43 or 45 vertices\\
Five-cycle blow-ups & five parts of 49, 51 or 53 vertices in a cycle, each part joined completely to its two neighbours\\
Cycle powers & odd cycles of 125--133 vertices, each vertex joined to the 32 nearest vertices on either side\\
\bottomrule
\end{tabularx}
\end{table}

\subsection{Suite B}
\label{sec:suite-b}

Suite~B has 50 graphs in 18 configurations (Table~\ref{tab:suite-b-design}), with three topology--weight seed pairs per configuration, except two ER configurations with two and the two-million-vertex ER configuration with one.

\begin{table}[htbp]\centering\small
\caption{Suite~B configurations and numbers of admitted graphs. Panels share baseline graphs, so their counts overlap.}
\label{tab:suite-b-design}
\begin{tabularx}{\linewidth}{@{}L{0.19\linewidth}L{0.53\linewidth}Y@{}}
\toprule
\textbf{Panel} & \textbf{Configurations} & \textbf{Graphs}\\\midrule
ER scalability & $n\in\{250000,500000,750000,1000000\}$; $d=32$ & 11\\
ER large limit & $n=2000000$; $d=32$; one seed & 1\\
ER density & $n\in\{500000,1000000\}$; $d\in\{32,64\}$ & 12\\
ER weight range & $n\in\{500000,1000000\}$; $d=32$; $W\in\{1000,10^6\}$ & 11\\
Bipartite ER & $n=500000$; $d=32$; equal sides & 3\\
SBM & $n=500000$; 100 communities; $d_{\mathrm{in}}/d_{\mathrm{out}}=28/4$ & 3\\
Toroidal geometric & $n=500000$; expected $d=32$ & 3\\
Watts--Strogatz & $n=500000$; $d=32$; $\beta=0.05$ & 3\\
ER / clique blocks & $n=512000$; ER $d=32$ versus $16000K_{32}$ & 6\\
ER / clique blocks & $n=1024000$; ER $d=32$ versus $32000K_{32}$ & 6\\
Biclique blocks & $n=1024000$; $16000K_{32,32}$ & 3\\\bottomrule
\end{tabularx}\end{table}

Except in the weight panel, $W=1000$. The ER configuration of each order serves as the baseline of several panels, and the 500,000-vertex ER graphs also anchor the structural comparison.

Weights are independent uniform integers in $\{1,\ldots,W\}$. The weight panel reuses topology seeds across weight ranges. The clique and biclique block graphs have a fixed structure, so their seeds differ only in the weights. Stochastic models use separate random streams for topology and weights. ER graphs, in the form introduced by Gilbert~\cite{Gilbert1959,ErdosRenyi1959}, include each unordered pair independently with probability $p=d/(n-1)$; bipartite ER graphs include each pair across two equal sides with probability $p=d/(n/2)$.

In the stochastic block model (SBM)~\cite{HollandLaskeyLeinhardt1983}, the 500,000 vertices form 100 equal communities, with within- and between-community edge probabilities $28/4999$ and $4/495000$, so the expected degree is $28+4=32$. Geometric graphs place the vertices uniformly on the unit torus and join two vertices at distance at most $\sqrt{32/((n-1)\pi)}$, where $\pi$ is the circle constant, which gives expected degree~32. Watts--Strogatz graphs~\cite{WattsStrogatz1998} start from a ring in which each vertex is joined to its 16 nearest neighbours on each side; each edge is rewired once with probability~0.05 to a uniformly chosen endpoint, avoiding loops and duplicates. Rewiring keeps $m=8{,}000{,}000$ and the mean degree~32, but not the degree of every vertex.

ER graphs and the stochastic structural graphs are conditioned on connectivity, with at most eight deterministic retries. Block graphs are disconnected by design. A $K_{32}$ block has degree~31, so its density only approximates that of the ER graphs with expected degree~32. With positive weights, every \MWM{} of a graph made of even cliques or of balanced complete bipartite blocks is perfect: a nonperfect matching leaves two adjacent free vertices, and adding their edge increases the weight. The stochastic families have no such guarantee and need the oracle screen.

\subsection{Platform}
\label{sec:platform}

All three programs are built from the same sources with GCC~11.4.1 as C++17 at optimization level \texttt{-O3} for the x86-64-v3 instruction set, with assertions, profiling and diagnostic output disabled (Appendix~\ref{app:campaign-evidence} lists the flags); one binary runs both suites and reads the SuiteSparse and generated files of Suite~A with a reader that accepts their formats. They run on exclusive nodes with two AMD EPYC~9334 processors (32 cores each, two NUMA nodes) under Linux~5.14, one program at a time, each in a fresh process on one CPU with one OpenMP thread. Each process has a 15-hour limit within an allocation of 240\,GiB. In each trial, Suite~A runs \Edm{} and \Gab{} back to back in an order that alternates across trials, and Suite~B runs the three programs in an order that rotates across trials. All nodes have this configuration.

\subsection{Timing and memory}
\label{sec:measurement}

Each graph--program pair has three trials in separate processes, without warm-up runs; six Suite~A graphs (\texttt{erb17}, \texttt{erb25}, \texttt{erb26}, \texttt{europe\_osm}, \texttt{kmer\_V2a} and \texttt{nlpkkt200}) have six. A table entry is the median over the trials, and we report trial-to-trial and seed-to-seed variation separately. Figures~\ref{fig:rq1} and~\ref{fig:stale} and the averages in Section~\ref{sec:results} compare \Gab{} with \Edm{} through the time ratio
\begin{equation}
 \operatorname*{median}_{k}\frac{t_{\mathrm{Gabow},k}}{t_{\mathrm{Edmonds},k}},
\end{equation}
where $t_{P,k}$ is the time of program $P$ in trial~$k$. The programs of a trial run back to back on one node, so each per-trial ratio compares them under the same conditions. The ratio exceeds~1 when \Edm{} is faster; it is the speedup of \Edm{} over \Gab{}, and its inverse is the speedup of \Gab{} over \Edm{}. The ratio of \Scal{} to \Edm{}, in Figure~\ref{fig:rq2}, is defined in the same way, and the speedup of \Scal{} is its inverse. Because these ratios are medians of per-trial ratios, they can differ slightly from the ratios of the median times in Tables~\ref{tab:suite-a-results} and~\ref{tab:suite-b-results}. A memory ratio is the ratio of the largest peaks over the trials. Summaries over several graphs are geometric means.

The memory of a program is the peak resident set size of its process, as reported by the operating system through \texttt{getrusage}; it covers the whole process, including the input graph. A table entry is the largest peak over the trials.

\section{Results}
\label{sec:results}

Tables~\ref{tab:suite-a-results} and~\ref{tab:suite-b-results} list every graph of the two suites, with the median times in seconds, the shortest time of each graph in bold, and the largest peak memory over the trials in GiB. Table~\ref{tab:suite-a-results} groups its rows as SuiteSparse biadjacency, generated random, generated random bipartite and SuiteSparse symmetric graphs, followed by the structured graphs in three groups: bipartite blocks, intervals and neighbourhoods, and odd-cycle blocks; the number after a structured graph is its topology seed. In Table~\ref{tab:suite-b-results}, a configuration gives the family, the order, the expected degree or block size and the largest weight, and its seeds~1--3 are separately generated graphs.

\subsection{Outcomes}

Every trial of every program completed and passed validation: 330 solves in Suite~A and 450 in Suite~B. The longest took 9{,}974\,s (\Gab{} on \texttt{er1}), and the largest peak memory was 136.9\,GiB (\Gab{} on \texttt{ab305}). The slowest and fastest trials of a program on a graph differ by a median of 0.7\% and at most 7.3\% in Suite~B, and by a median of 0.4\% and at most 1.4\% on the structured graphs of Suite~A. On some other large graphs of Suite~A they differ by up to 90\% (Section~\ref{sec:threats}).

\begingroup\footnotesize\setlength{\tabcolsep}{2.2pt}
\begin{longtable}{@{}l r r r r r r r@{}}
\caption{Suite~A: median solver times and peak memory. The shorter time is in bold.}\label{tab:suite-a-results}\\
\toprule
 & & & & \multicolumn{2}{c}{\textbf{Median time (s)}} & \multicolumn{2}{c}{\textbf{Peak memory (GiB)}}\\\cmidrule(lr){5-6}\cmidrule(l){7-8}
\textbf{Graph} & $n$ & $m$ & Avg.\ deg. & \Edm{} & \Gab{} & \Edm{} & \Gab{}\\\midrule
\endfirsthead
\toprule
 & & & & \multicolumn{2}{c}{\textbf{Median time (s)}} & \multicolumn{2}{c}{\textbf{Peak memory (GiB)}}\\\cmidrule(lr){5-6}\cmidrule(l){7-8}
\textbf{Graph} & $n$ & $m$ & Avg.\ deg. & \Edm{} & \Gab{} & \Edm{} & \Gab{}\\\midrule
\endhead
\bottomrule
\endfoot
GL7d19 & 3,866,439 & 37,322,725 & 19.31 & \textbf{1706.6} & 1835.6 & 17.3 & 29.4\\
Hardesty3 & 15,809,384 & 40,451,632 & 5.12 & \textbf{272.6} & 301.8 & 29.5 & 45.7\\
relat9 & 12,909,396 & 38,955,420 & 6.04 & \textbf{89.7} & 105.1 & 26.1 & 41.6\\
spal\_004 & 331,899 & 46,168,124 & 278.21 & 63.4 & \textbf{62.3} & 13.8 & 26.6\\
\midrule
ab105 & 10,000,000 & 49,999,972 & 10.00 & \textbf{442.9} & 516.0 & 25.8 & 40.3\\
ab205 & 20,000,000 & 99,999,969 & 10.00 & \textbf{985.8} & 1144.8 & 59.6 & 83.6\\
ab305 & 30,000,000 & 149,999,971 & 10.00 & \textbf{1580.9} & 1932.3 & 78.7 & 136.9\\
er1 & 1,000,000 & 49,996,668 & 99.99 & \textbf{9777.7} & 9960.5 & 17.8 & 28.3\\
\midrule
erb7 & 1,100,000 & 88,811,244 & 161.47 & \textbf{264.8} & 291.6 & 27.3 & 52.9\\
erb8 & 1,600,000 & 133,196,375 & 166.50 & \textbf{355.3} & 439.9 & 38.8 & 65.0\\
erb9 & 2,100,000 & 177,598,334 & 169.14 & \textbf{448.8} & 453.5 & 54.4 & 105.6\\
erb16 & 1,150,000 & 133,227,024 & 231.70 & \textbf{561.4} & 688.1 & 39.6 & 65.7\\
erb17 & 1,650,000 & 199,828,098 & 242.22 & \textbf{737.4} & 918.1 & 59.4 & 109.8\\
erb18 & 2,150,000 & 266,444,173 & 247.86 & \textbf{1351.9} & 1878.6 & 76.5 & 128.5\\
erb25 & 1,200,000 & 177,624,362 & 296.04 & \textbf{951.0} & 1005.9 & 53.4 & 104.4\\
erb26 & 1,700,000 & 266,406,205 & 313.42 & \textbf{1386.6} & 1767.6 & 76.5 & 128.5\\
\midrule
asia\_osm & 11,950,757 & 12,711,603 & 2.13 & \textbf{63.4} & 79.4 & 21.9 & 27.1\\
com-LiveJournal & 3,997,962 & 34,681,189 & 17.35 & \textbf{403.4} & 429.1 & 16.1 & 29.0\\
europe\_osm & 50,912,018 & 54,054,660 & 2.12 & \textbf{313.7} & 393.3 & 93.2 & 115.2\\
hugebubbles-00010 & 19,458,087 & 29,179,764 & 3.00 & \textbf{195.5} & 255.0 & 38.0 & 48.0\\
kmer\_V2a & 55,042,369 & 58,608,800 & 2.13 & \textbf{414.4} & 561.7 & 101.0 & 124.9\\
kron\_g500-logn21 & 2,097,152 & 91,040,932 & 86.82 & \textbf{1043.0} & 1150.2 & 28.9 & 54.6\\
mtx2010 & 914,231 & 2,228,136 & 4.87 & \textbf{11.4} & 13.7 & 2.1 & 2.8\\
nlpkkt200 & 16,240,000 & 215,992,816 & 26.60 & \textbf{5451.0} & 6785.8 & 89.3 & 134.8\\
rgg\_n\_2\_23\_s0 & 8,388,608 & 63,501,393 & 15.14 & \textbf{566.8} & 612.7 & 31.3 & 45.3\\
\midrule
Biclique blocks (1) & 1,001,109 & 124,342,272 & 248.41 & 155.7 & \textbf{73.6} & 36.0 & 60.5\\
Biclique blocks (2) & 1,001,073 & 124,194,816 & 248.12 & 155.0 & \textbf{72.9} & 35.9 & 60.4\\
Connected blocks (1) & 1,001,109 & 124,342,272 & 248.41 & 155.6 & \textbf{73.0} & 36.0 & 60.5\\
Connected blocks (2) & 1,001,073 & 124,194,816 & 248.12 & 155.0 & \textbf{73.1} & 35.9 & 60.4\\
Biclique ring (1) & 1,001,109 & 124,342,272 & 248.41 & 156.6 & \textbf{73.8} & 36.0 & 60.5\\
Biclique ring (2) & 1,001,073 & 124,194,816 & 248.12 & 155.5 & \textbf{74.1} & 35.9 & 60.4\\
Ferrers blocks (1) & 1,001,109 & 93,243,545 & 186.28 & 132.1 & \textbf{68.2} & 28.3 & 53.7\\
Ferrers blocks (2) & 1,001,073 & 93,133,343 & 186.07 & 131.3 & \textbf{68.1} & 28.2 & 53.7\\
Convex windows (1) & 1,001,109 & 62,171,136 & 124.20 & 77.0 & \textbf{44.5} & 18.6 & 30.9\\
Convex windows (2) & 1,001,073 & 62,097,408 & 124.06 & 77.3 & \textbf{44.5} & 18.5 & 30.9\\
\midrule
Circular intervals (1) & 999,999 & 124,121,088 & 248.24 & 142.5 & \textbf{88.7} & 35.9 & 60.4\\
Circular intervals (2) & 999,999 & 124,121,088 & 248.24 & 142.9 & \textbf{88.7} & 35.9 & 60.4\\
Cube neighbourhoods (1) & 999,565 & 100,239,550 & 200.57 & 130.4 & \textbf{91.5} & 30.0 & 55.2\\
Cube neighbourhoods (2) & 999,041 & 100,066,106 & 200.32 & 154.5 & \textbf{91.8} & 29.9 & 55.2\\
\midrule
Odd cliques (1) & 999,740 & 63,996,378 & 128.03 & 223.9 & \textbf{194.7} & 21.9 & 32.7\\
Odd cliques (2) & 1,000,232 & 64,060,532 & 128.09 & 224.0 & \textbf{193.1} & 21.9 & 32.7\\
Clique chain (1) & 999,740 & 64,004,129 & 128.04 & 225.9 & \textbf{195.1} & 21.9 & 32.7\\
Clique chain (2) & 1,000,232 & 64,068,283 & 128.11 & 225.9 & \textbf{194.2} & 21.9 & 32.7\\
Tripartite (1) & 999,180 & 42,991,152 & 86.05 & 173.2 & \textbf{162.0} & 16.9 & 26.8\\
Tripartite (2) & 999,732 & 43,037,880 & 86.10 & 172.0 & \textbf{161.1} & 16.5 & 26.8\\
Five-cycle blow-ups (1) & 1,000,330 & 51,080,730 & 102.13 & 230.4 & \textbf{215.8} & 18.6 & 28.5\\
Five-cycle blow-ups (2) & 999,950 & 51,041,770 & 102.09 & 230.8 & \textbf{215.5} & 18.6 & 28.5\\
Cycle powers (1) & 999,740 & 31,991,680 & 64.00 & 138.4 & \textbf{132.7} & 11.5 & 17.1\\
Cycle powers (2) & 1,000,232 & 32,007,424 & 64.00 & 138.4 & \textbf{132.2} & 11.5 & 17.1\\
\end{longtable}\endgroup

\begingroup\footnotesize\setlength{\tabcolsep}{2.2pt}
\begin{longtable}{@{}l c r r r r r r@{}}
\caption{Suite~B: median solver times and peak memory. The shortest time is in bold.}\label{tab:suite-b-results}\\
\toprule
 & & \multicolumn{3}{c}{\textbf{Median time (s)}} & \multicolumn{3}{c}{\textbf{Peak memory (GiB)}}\\\cmidrule(lr){3-5}\cmidrule(l){6-8}
\textbf{Configuration} & \textbf{Seed} & \Edm{} & \Gab{} & \Scal{} & \Edm{} & \Gab{} & \Scal{}\\\midrule
\endfirsthead
\toprule
 & & \multicolumn{3}{c}{\textbf{Median time (s)}} & \multicolumn{3}{c}{\textbf{Peak memory (GiB)}}\\\cmidrule(lr){3-5}\cmidrule(l){6-8}
\textbf{Configuration} & \textbf{Seed} & \Edm{} & \Gab{} & \Scal{} & \Edm{} & \Gab{} & \Scal{}\\\midrule
\endhead
\bottomrule
\endfoot
ER 250k d32 W1k & 1 & 210.0 & 224.9 & \textbf{38.8} & 2.2 & 3.0 & 1.3\\
ER 250k d32 W1k & 2 & 197.3 & 208.7 & \textbf{37.5} & 2.2 & 3.1 & 1.3\\
ER 250k d32 W1k & 3 & 304.0 & 354.9 & \textbf{38.2} & 2.3 & 3.1 & 1.3\\
ER 500k d32 W1k & 1 & 623.4 & 708.7 & \textbf{118.4} & 4.6 & 6.3 & 2.6\\
ER 500k d32 W1k & 2 & 511.3 & 532.0 & \textbf{120.9} & 4.6 & 6.3 & 2.6\\
ER 500k d32 W1k & 3 & 475.8 & 500.4 & \textbf{114.1} & 4.6 & 6.3 & 2.6\\
ER 500k d32 W1M & 1 & 471.7 & 488.3 & \textbf{129.3} & 4.6 & 6.2 & 2.6\\
ER 500k d32 W1M & 2 & 448.5 & 469.6 & \textbf{128.6} & 4.4 & 6.1 & 2.6\\
ER 500k d32 W1M & 3 & 562.1 & 631.7 & \textbf{129.6} & 4.6 & 6.2 & 2.6\\
ER 500k d64 W1k & 1 & 1493.6 & 1521.9 & \textbf{188.6} & 8.5 & 11.6 & 4.1\\
ER 500k d64 W1k & 2 & 1544.9 & 1593.0 & \textbf{193.9} & 8.5 & 11.5 & 4.1\\
ER 500k d64 W1k & 3 & 1648.2 & 1680.8 & \textbf{199.9} & 8.6 & 11.7 & 4.1\\
ER 512k d32 W1k & 1 & 606.4 & 629.8 & \textbf{128.9} & 4.9 & 6.6 & 2.7\\
ER 512k d32 W1k & 2 & 571.4 & 596.6 & \textbf{119.7} & 4.8 & 6.5 & 2.7\\
ER 512k d32 W1k & 3 & 503.0 & 522.4 & \textbf{118.3} & 4.7 & 6.4 & 2.7\\
ER 750k d32 W1k & 2 & 861.5 & 893.0 & \textbf{272.9} & 7.1 & 9.6 & 3.7\\
ER 750k d32 W1k & 3 & 3091.9 & 3518.7 & \textbf{263.1} & 7.4 & 9.9 & 3.7\\
ER 1000k d32 W1k & 1 & 1381.5 & 1506.5 & \textbf{434.8} & 9.7 & 13.0 & 5.1\\
ER 1000k d32 W1k & 2 & 2582.9 & 2958.3 & \textbf{454.9} & 9.8 & 13.1 & 5.1\\
ER 1000k d32 W1k & 3 & 1254.6 & 1295.5 & \textbf{443.9} & 9.7 & 13.0 & 5.1\\
ER 1000k d32 W1M & 2 & 1427.5 & 1474.5 & \textbf{489.8} & 9.5 & 12.8 & 5.1\\
ER 1000k d32 W1M & 3 & 1255.7 & 1362.4 & \textbf{479.6} & 9.3 & 12.6 & 5.1\\
ER 1000k d64 W1k & 1 & 6711.4 & 7161.6 & \textbf{974.8} & 19.1 & 25.2 & 8.1\\
ER 1000k d64 W1k & 2 & 5759.7 & 6118.9 & \textbf{977.4} & 18.8 & 24.9 & 8.1\\
ER 1000k d64 W1k & 3 & 4772.3 & 4938.7 & \textbf{993.2} & 18.6 & 24.8 & 8.1\\
ER 1024k d32 W1k & 1 & 2118.1 & 2526.2 & \textbf{434.4} & 10.4 & 13.8 & 5.3\\
ER 1024k d32 W1k & 2 & 1398.6 & 1451.1 & \textbf{466.8} & 10.0 & 13.4 & 5.3\\
ER 1024k d32 W1k & 3 & 4007.8 & 4567.8 & \textbf{452.9} & 10.4 & 13.8 & 5.3\\
ER 2000k d32 W1k & 1 & 3814.9 & 3918.2 & \textbf{1177.7} & 20.9 & 27.5 & 10.2\\
Bip. ER 500k d32 W1k & 1 & 477.0 & 496.1 & \textbf{99.9} & 4.4 & 6.0 & 2.4\\
Bip. ER 500k d32 W1k & 2 & 500.1 & 514.2 & \textbf{112.2} & 4.4 & 6.1 & 2.4\\
Bip. ER 500k d32 W1k & 3 & 475.2 & 491.8 & \textbf{95.5} & 4.4 & 6.1 & 2.4\\
SBM 500k 28/4 W1k & 1 & 616.6 & 704.1 & \textbf{84.5} & 4.6 & 6.2 & 2.6\\
SBM 500k 28/4 W1k & 2 & 477.1 & 494.4 & \textbf{88.8} & 4.6 & 6.3 & 2.6\\
SBM 500k 28/4 W1k & 3 & 504.5 & 521.3 & \textbf{87.5} & 4.7 & 6.3 & 2.6\\
Torus 500k d32 W1k & 1 & \textbf{70.1} & 86.0 & 2038.1 & 3.2 & 4.7 & 44.2\\
Torus 500k d32 W1k & 2 & \textbf{67.6} & 81.8 & 1758.5 & 3.2 & 4.6 & 36.9\\
Torus 500k d32 W1k & 3 & \textbf{70.0} & 85.7 & 2289.8 & 3.2 & 4.7 & 71.0\\
WS 500k d32 b0.05 W1k & 1 & 121.8 & 153.6 & \textbf{79.7} & 3.2 & 4.8 & 2.7\\
WS 500k d32 b0.05 W1k & 2 & \textbf{74.6} & 83.5 & 80.9 & 3.2 & 4.8 & 2.7\\
WS 500k d32 b0.05 W1k & 3 & \textbf{73.8} & 81.0 & 80.7 & 3.2 & 4.8 & 2.7\\
Clique 512k K32 W1k & 1 & \textbf{15.8} & 19.7 & 25.6 & 3.2 & 4.6 & 2.6\\
Clique 512k K32 W1k & 2 & \textbf{16.1} & 19.9 & 25.5 & 3.2 & 4.7 & 2.6\\
Clique 512k K32 W1k & 3 & \textbf{16.2} & 20.0 & 25.4 & 3.2 & 4.7 & 2.6\\
Clique 1024k K32 W1k & 1 & \textbf{40.2} & 42.5 & 53.6 & 6.3 & 9.2 & 5.2\\
Clique 1024k K32 W1k & 2 & \textbf{39.2} & 47.2 & 54.4 & 6.3 & 9.2 & 5.2\\
Clique 1024k K32 W1k & 3 & \textbf{34.3} & 42.0 & 53.9 & 6.3 & 9.2 & 5.2\\
Biclique 1024k K32,32 W1k & 1 & \textbf{35.5} & 42.5 & 45.1 & 6.4 & 9.4 & 5.0\\
Biclique 1024k K32,32 W1k & 2 & \textbf{35.4} & 42.3 & 45.9 & 6.4 & 9.4 & 5.0\\
Biclique 1024k K32,32 W1k & 3 & \textbf{35.1} & 42.6 & 45.7 & 6.4 & 9.4 & 5.0\\
\end{longtable}\endgroup

\subsection{RQ1: Gabow's data structures}

Gabow's candidate structure made \Gab{} faster on all 24 structured graphs and on one SuiteSparse graph. Because \Edm{} and \Gab{} perform the same search (Section~\ref{sec:validation}), the time ratio of \Gab{} measures the cost of its candidate management. We summarize the structured graphs of Suite~A separately from its other graphs, because they were chosen to favour \Gab{} (Section~\ref{sec:suite-a}). On the other 25 graphs of Suite~A, \Gab{} took longer than \Edm{} on 24, by 1.0--38.8\%, and was 1.8\% faster on \texttt{spal\_004}, in each of its three trials; on average, \Edm{} was 1.16 times as fast as \Gab{}. In Suite~B, \Gab{} took longer on all 50 graphs, by 1.9--26.1\%, and \Edm{} was on average 1.10 times as fast. On the structured graphs, \Gab{} was faster on all 24 and on average 1.50 times as fast as \Edm{}: faster by 42--53\% on the five constructions from bipartite blocks, by 30--41\% on circular intervals and cube neighbourhoods, and by 4--14\% on the five constructions from odd-cycle blocks. On these graphs \Edm{} discards far more stale queue entries per search step than on the others. In the bipartite constructions nearly every edge enters its queue as a candidate and becomes stale once an endpoint is matched, so \Edm{} pays a queue operation for each of 62--124 million entries, 997--2,006 per search step, whereas \Gab{} removes such records when their trees retire and never selects them; on the odd-cycle constructions \Edm{} discards 30--59 entries per search step. Selecting candidates took 36--79\% of the time of \Edm{} on these graphs (Section~\ref{sec:profiles} and Figure~\ref{fig:profiles}). These counts do not explain the whole difference: the random bipartite graphs of Suite~A have similar counts, yet \Gab{} was slower on them. One difference is the vertex numbering: in the structured graphs, adjacent vertices have nearby numbers, so the data that the programs keep for them lies close together in memory. This may help \Gab{} more than \Edm{}, because \Gab{} keeps more data (Section~\ref{sec:threats}).

\begin{figure}[tbp]
  \centering
  \includegraphics[width=\linewidth]{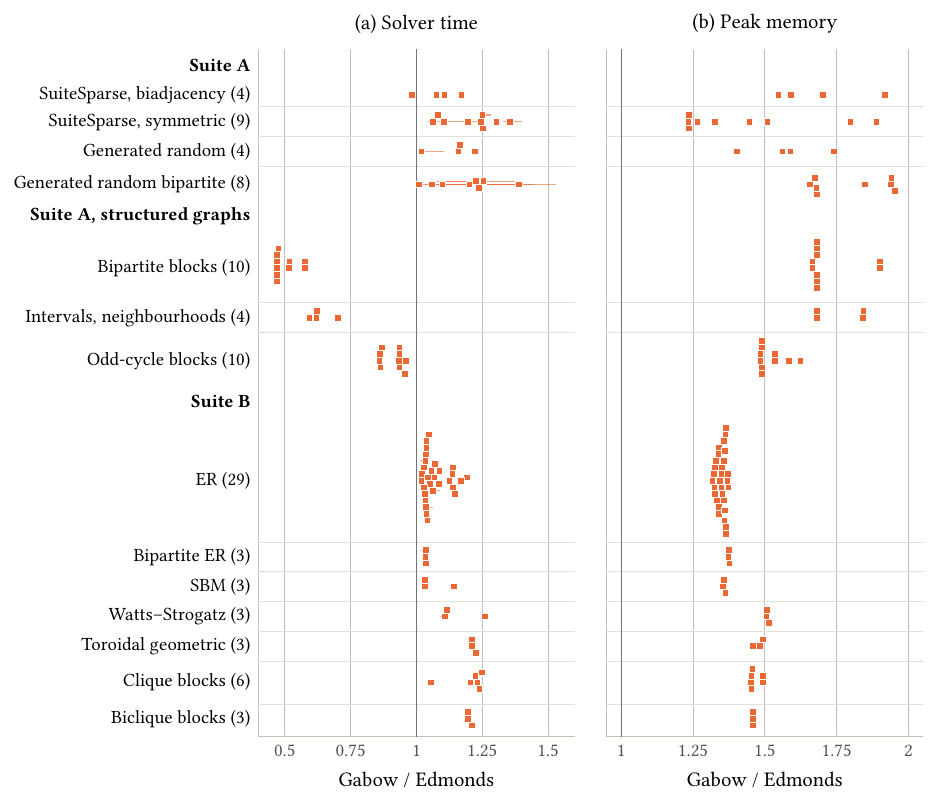}
  \caption{Solver time and peak memory of \Gab{} relative to \Edm{}.}
  \Description{Two strip plots with one row per suite and family and one marker per graph. In panel a, the time ratios of Gabow to Edmonds lie between 0.47 and 0.96 on the structured graphs of Suite A and between 1 and 1.4 on all other graphs, except one SuiteSparse graph at 0.98. In panel b, all memory ratios lie between 1.23 and 1.95.}
  \label{fig:rq1}
\end{figure}

Figure~\ref{fig:rq1} shows the ratios by suite and family, one marker per graph; a horizontal line gives the range of the per-trial time ratios where it is wider than the marker, and values above~1 mean that \Gab{} needs more. In Suite~B the geometric mean is smallest on bipartite ER and SBM graphs, 1.036 and 1.067, and largest on clique blocks, biclique blocks and toroidal geometric graphs, 1.199, 1.201 and 1.216. On the 29 ER graphs the ratio rises with the number of contractions per vertex, with Spearman's rank correlation 0.76, but not with the number of expansions per vertex ($-0.04$). On the structured graphs the two seeds of a construction give ratios within 0.01 of each other, except for the cube neighbourhoods, 0.70 and 0.59, where \Edm{} took 130 and 154\,s and \Gab{} 92\,s on both. The ratios of the other Suite~A graphs vary more, because three nodes slowed runs with large memory peaks, \Gab{} more than \Edm{}: on \texttt{erb18}, the per-trial ratio was 1.39 in the two trials on such a node and 1.09 in the third (Section~\ref{sec:threats}). \Gab{} also needed more memory on every graph, 1.23--1.95 times as much as \Edm{} in Suite~A, of which 1.49--1.90 on its structured graphs, and 1.32--1.51 times in Suite~B.

\subsection{RQ2: Weight scaling}
\label{sec:rq2}

\Scal{} had the lowest median on 36 of the 50 Suite~B graphs and \Edm{} on the other 14. \Scal{} led on all 29 ER graphs, all three bipartite ER graphs, all three SBM graphs and one of the three Watts--Strogatz graphs; \Edm{} led on all three toroidal geometric graphs, all six clique-block graphs, all three biclique-block graphs and the other two Watts--Strogatz graphs. On ER graphs \Scal{} was 2.6--11.8 times faster than \Edm{}, 4.83 times on average, and on bipartite ER and SBM graphs 4.4--5.0 and 5.4--7.3 times faster. On toroidal geometric graphs \Edm{} was 26.0--32.6 times faster than \Scal{}, and on clique and biclique blocks it was faster by factors of 1.33--1.62 and 1.27--1.30. Over the 50 graphs, \Scal{} was on average 2.35 times as fast as \Edm{} and 2.59 times as fast as \Gab{}; these averages reflect our inventory, in which 29 of the 50 graphs are ER graphs, and the ranges by family are more informative.

\begin{figure}[tbp]
  \centering
  \includegraphics[width=\linewidth]{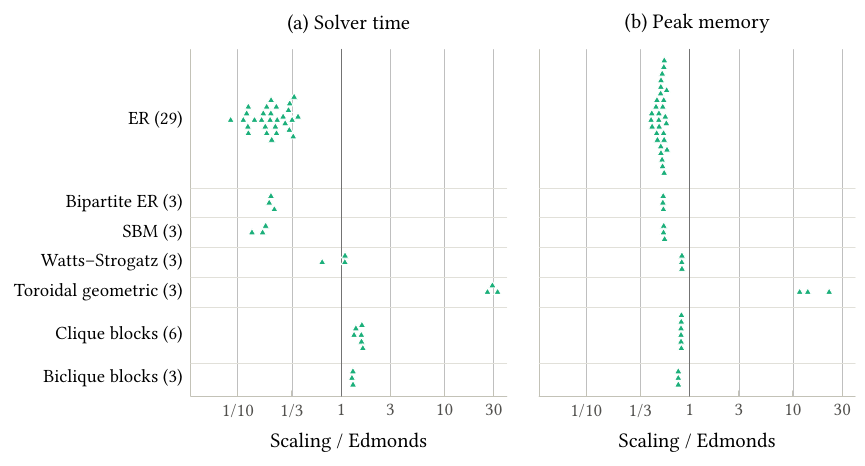}
  \caption{Solver time and peak memory of \Scal{} relative to \Edm{} on the 50 Suite~B graphs.}
  \Description{Two strip plots on logarithmic axes with one row per family and one marker per graph. In panel a, the time ratios of Scaling to Edmonds are below 1 on ER, bipartite ER and SBM graphs, and about 26 to 33 on toroidal geometric graphs. In panel b, the memory ratios are below 1 except on toroidal geometric graphs, where they are about 12 to 22.}
  \label{fig:rq2}
\end{figure}

Figure~\ref{fig:rq2} shows these ratios with the corresponding memory ratios on logarithmic axes; values below~1 mean that \Scal{} needs less. Except on Watts--Strogatz graphs, the ordering of \Scal{} and \Edm{} is the same for every graph of a family, and the per-trial ratios of a graph differ by a median of 1.1\% and at most 9.0\%, little compared with the differences between families.

The instrumented runs of Section~\ref{sec:profiles} explain most of these differences. On ER graphs, the times of \Edm{} and \Gab{} depend on whether a large blossom forms through many nested contractions, which varies between seeds. Each contraction lists the vertices of the new blossom (Section~\ref{sec:gabow-implementation}), so the largest speedup of \Scal{}, 11.8, occurs on the 750,000-vertex graph whose contractions list the most vertices among the instrumented graphs (Section~\ref{sec:profiles}). On the block graphs, the searches of \Edm{} and \Gab{} stay inside components of 32 or 64 vertices, whereas the Criterion-2 phases of \Scal{} scanned each edge about four times per scale, in each of 28 or 29 scales. On the profiled geometric graph, \Scal{} scanned each edge 49 times per scale, against 8.6 times on an ER graph of the same order, and spent nearly half of its time dismantling inherited blossoms nested tens of thousands of levels deep.

\subsection{RQ3: Order, density, weights and structure}

\begin{figure}[tbp]
  \centering
  \includegraphics[width=\linewidth]{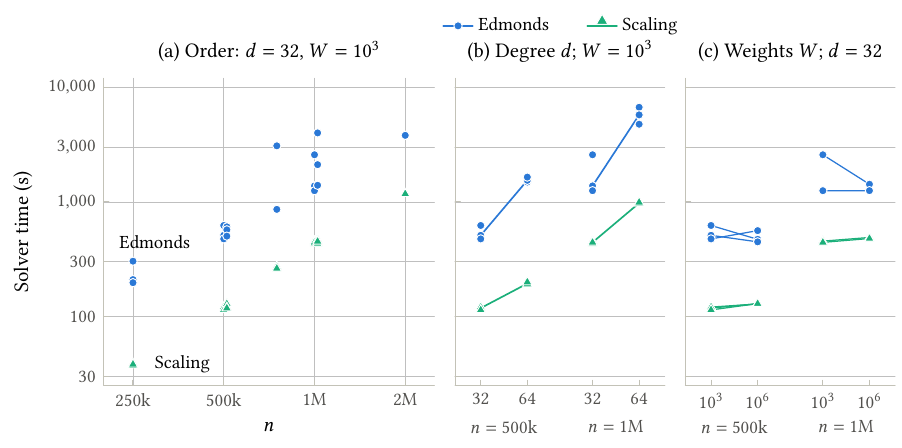}
  \caption{Median times of \Edm{} and \Scal{} on ER graphs by (a) order, (b) expected degree and (c) largest weight.}
  \Description{Three panels of median times on ER graphs. Panel a plots time against order, with Scaling below Edmonds at every order and Edmonds varying strongly between seeds. Panel b compares expected degrees 32 and 64. Panel c compares largest weights 1000 and one million, with Scaling nearly flat and Edmonds changing in both directions.}
  \label{fig:rq3}
\end{figure}

Figure~\ref{fig:rq3} shows the ER graphs, one marker per graph: in panel~(a) the 18 graphs with $d=32$ and $W=10^3$; in panel~(b) the expected degrees 32 and 64, with lines joining the medians over the seeds; and in panel~(c) the largest weights $10^3$ and $10^6$, with lines joining graphs of the same topology. \Gab{} is omitted; its times on these graphs are 2--19\% above those of \Edm{}. Running time grows with order for all three programs. From 250,000 to 2,000,000 vertices, the median time of \Scal{} grew by a factor of 30.8 and that of \Edm{} by 18.2, so the speedup of \Scal{} on the two-million-vertex graph, 3.2, is near the low end of its range on ER graphs; a single graph at this order does not determine a growth law. The times of \Edm{} and \Gab{} vary strongly between the seeds of one configuration, by factors of up to 1.5 at 250,000 vertices, 2.9 at 1,024,000 and 3.6 at 750,000, whereas those of \Scal{} vary by at most 9\%. The number of search steps varies much less: on the slower of the two 750,000-vertex graphs, the main loop of \Edm{} ran 15\% more iterations than on the faster one but took 3.6 times as long, so each iteration cost more; Section~\ref{sec:profiles} relates the difference to blossom contraction.

Doubling the expected degree from 32 to 64 multiplied the median time of \Scal{} by 1.6 at 500,000 vertices and 2.2 at one million, and that of \Edm{} by 3.0 and 4.2. Raising the largest weight from $W=10^3$ to $W=10^6$ on the same topologies adds ten scales, from 28 to 38 at 500,000 vertices and from 29 to 39 at one million, and increased the time of \Scal{} by 6--13\% on all five topologies. The times of \Edm{} and \Gab{} changed in both directions, by factors from 0.55 to 1.18 for \Edm{}, comparable to their variation between seeds.

The structural panels show the clearest reversals. \Scal{} leads on the ER graphs at both block orders and on bipartite ER and SBM graphs, whereas \Edm{} leads on geometric graphs and on clique and biclique blocks; the Watts--Strogatz graphs are mixed across seeds. Odd cycles alone do not explain these differences: connectivity, component size and density change together, and the two bipartite families favour different programs.

\subsection{Memory}

Peak memory covers the whole process, including the input graph. In Suite~A, the peak of \Edm{} ranged from 2.1\,GiB on \texttt{mtx2010} to 101.0\,GiB on \texttt{kmer\_V2a}, and \Gab{} needed 1.23--1.95 times as much; its 136.9\,GiB on \texttt{ab305} is the largest peak of the study. On the structured graphs, the peak of \Edm{} ranged from 11.5 to 36.0\,GiB, and \Gab{} needed 1.49--1.90 times as much. In Suite~B, \Scal{} needed less memory than \Edm{} on 47 of the 50 graphs, 0.42--0.84 times as much, but 11.6--22.4 times as much on the three toroidal geometric graphs, where its peak reached 36.9--71.0\,GiB against 3.2\,GiB for \Edm{} (Figure~\ref{fig:rq2}b); Section~\ref{sec:profiles} traces this memory to deeply nested inherited blossoms.

\subsection{Phase times and operation counts}
\label{sec:profiles}
\label{sec:instrumented}

Three kinds of instrumented runs show where the work goes. Untimed instrumented runs of \Edm{} and \Gab{}, one pair per graph on all 99 graphs, count their search steps and the entries that \Edm{} inserts into its queue and later discards as stale (Section~\ref{sec:validation}). Counting builds recorded operation counts without clocks, once per graph and program, on 21 of the graphs: 17 of the 50 Suite~B graphs, from every family, including both Watts--Strogatz seeds whose rankings differ, two seeds each at 250,000, 750,000 and 1,024,000 vertices, both weight ranges and both densities; and the Suite~A graphs \texttt{spal\_004}, \texttt{erb16}, \texttt{rgg\_n\_2\_23\_s0} and \texttt{kron\_g500-logn21}. \Scal{} was also counted on all 50 Suite~B graphs. Profiling builds of the measured programs time exclusive phases on these 21 graphs and on the 24 structured graphs, in three trials per graph and program, with \Scal{} on the 17 Suite~B graphs among them; the phases cover at least 97.8\% of the solver time, and a phase time is the median over the trials. Neither the instrumentation nor the repairs of the measured build change the search: the numbers of contractions, expansions, additions to trees and tree teardowns in the profiled runs, and the vertices that the contractions list, equal those of the counting runs. The first profiled trial of each graph and program was paired with an uninstrumented run on the same node; both returned the same matching, and the profiled run took 1.00--1.05 times as long for \Edm{} and \Gab{} and 1.03--1.07 times for \Scal{}. Of the 107 uninstrumented runs, 103 took within 3\% of the medians of the main campaign; the other four, runs of \Gab{} with a peak memory of about 60\,GiB or more, took 0.86--1.10 times as long. All 428 profiling runs completed and returned the optimum weights of the main campaign. Six further runs of \Gab{}, three on the profiled geometric graph and three on an odd-clique graph, timed every NCA query.

\begin{figure}[tbp]
  \centering
  \includegraphics[width=\linewidth]{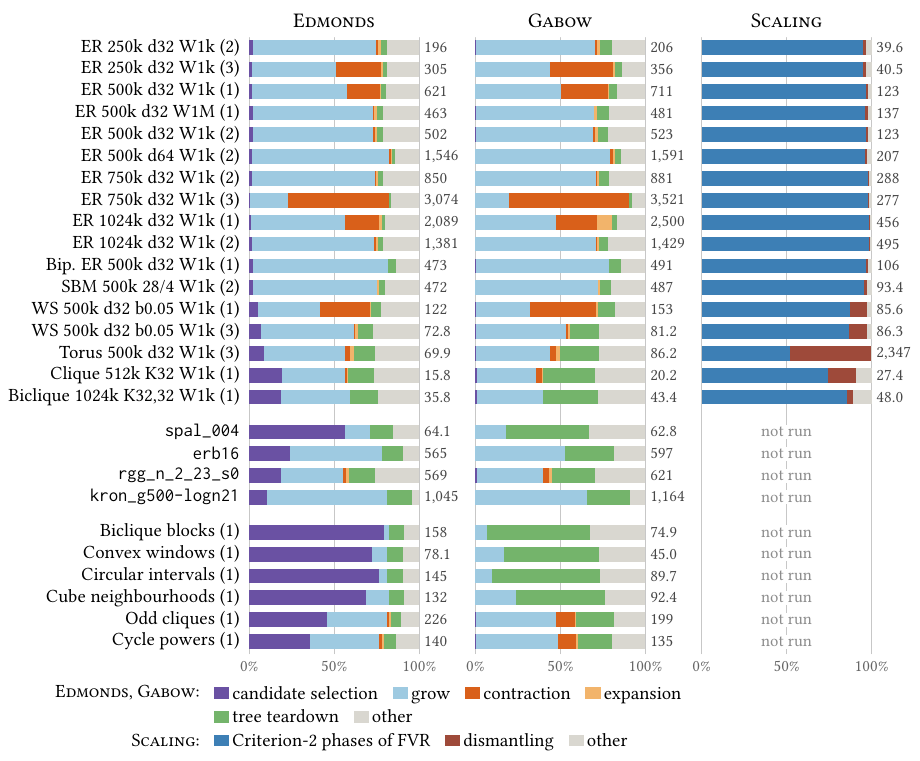}
  \caption{Division of solver time among phases on the 21 instrumented graphs and on six structured graphs; medians of three trials, phase by phase. Numbers give the totals in seconds.}
  \Description{Stacked horizontal bars with one row per instrumented graph, for Edmonds, Gabow and Scaling, with total seconds at the right. Contraction takes a large share of the time of Edmonds and Gabow on the slow ER seeds, tree teardown takes a larger share for Gabow than for Edmonds on every graph, and Scaling spends most of its time in the Criterion-2 phases except on the toroidal graph, where dismantling takes about half.}
  \label{fig:profiles}
\end{figure}

Figure~\ref{fig:profiles} shows how the three programs divide their time among phases on the 21 instrumented graphs and how \Edm{} and \Gab{} divide it on six of the structured graphs, two from each group; Tables~\ref{tab:profile-edmonds}--\ref{tab:profile-scaling} in Appendix~\ref{app:profiles} give the times, for \Edm{} and \Gab{} on all 24 structured graphs. The profiled runs confirm that \Edm{} and \Gab{} perform the same search: their numbers of contractions, expansions, additions to trees and tree teardowns, and of the vertices that contractions and teardowns touch, agree on all 45 graphs (Table~\ref{tab:counts}). For \Scal{}, the figure shows dismantling and the Criterion-2 phases separately and its other stages together, because none of those took more than 4.5\% of the time on any graph; dividing the Criterion-2 phases further would need timers that separate their operations, which these runs did not record.

\emph{Candidate management.} The two structures pay at different times. \Edm{} inserts candidates cheaply and pays only when a stale entry reaches the top of its queue: selecting the next candidate took it 3.1--134.6\,s, mostly for discarding stale entries, against at most 5.1\,s for \Gab{}. \Gab{} instead keeps its records current at every change of the search forest, and it took longer than \Edm{} for initialization, growing trees and tearing them down on all 21 graphs. Its initialization, in which both programs insert every edge as a candidate, took 1.6--2.2 times as long, and tearing down trees, which retires the records of their vertices, took 1.7--3.7 times as long and was its largest extra cost on 16 of the 21 graphs. On the ER graphs, candidate selection took at most 2.2\% of the time of \Edm{}, so the part of the work that Gabow's structure speeds up is small there. On \texttt{spal\_004} it took 56\%, because of 46.2 million stale entries, and the saving outweighed the extra costs, which is why \Gab{} was faster on this graph. NCA queries and their maintenance were not a dominant cost on the two graphs where they were timed: they took 2.3\% of the time on the profiled geometric graph and 4.1\% on an odd-clique graph, which made 47 million NCA queries, the most of the 45 profiled graphs.

\begin{figure}[tbp]
  \centering
  \includegraphics[width=0.82\linewidth]{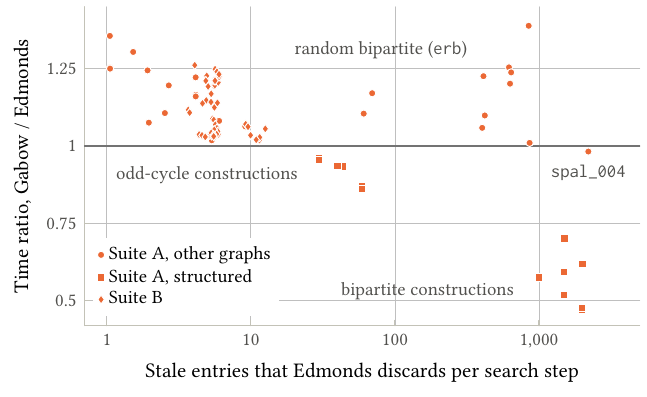}
  \caption{Time ratio of \Gab{} to \Edm{} against the stale entries that \Edm{} discards per search step; one marker per graph.}
  \Description{A scatter plot of the time ratio of Gabow to Edmonds against the number of stale queue entries that Edmonds discards per search step, on a logarithmic axis from 1 to 2,200, with one marker per graph and one marker shape for each of three groups: the structured graphs of Suite A, its other graphs, and Suite B. Suite B lies at 4 to 13 entries per step with ratios from 1.02 to 1.26. The structured graphs lie below 1: the bipartite constructions at 1,000 to 2,000 entries per step with ratios from 0.47 to 0.70, and the odd-cycle constructions at 30 to 59 entries per step with ratios from 0.86 to 0.96. Among the other graphs of Suite A, the random bipartite graphs lie at 400 to 860 entries per step with ratios from 1.01 to 1.39, and spal_004 at 2,200 entries per step with ratio 0.98.}
  \label{fig:stale}
\end{figure}

\emph{Stale entries.} Figure~\ref{fig:stale} plots the time ratio against the number of stale entries that \Edm{} discards per search step, where a search step is a growth, contraction, expansion or augmentation. In Suite~B this number is 3.7--12.6, and on the SuiteSparse and generated graphs of Suite~A it is 1.0--69, except on \texttt{spal\_004} and the eight \texttt{erb} graphs. On the bipartite structured graphs, every edge is a candidate between two trees at the start, and its entry becomes stale once an endpoint is matched: \Edm{} discards one entry per edge, 62--124 million in all, or 997--2,006 per search step. On the odd-cycle constructions it discards about two entries per edge, 30--59 per search step. \Gab{} was faster on all of these graphs. Candidate selection took 68--79\% of the time of \Edm{} on the bipartite constructions and 36--46\% on the odd-cycle ones; \Gab{} spent at most 0.7\,s on it, and its largest extra cost, tree teardown, took 2.7--4.2 times as long as that of \Edm{} (Figure~\ref{fig:profiles}). Many stale entries are not sufficient, however. The eight random bipartite graphs of Suite~A also have about one stale entry per edge, 403--861 per search step, and \Gab{} was slower on all of them, by 1--39\%, or by 1--12\% in the trials on the nodes that Section~\ref{sec:threats} finds unaffected; on \texttt{spal\_004}, with 2,200 per search step, it was 1.8\% faster. On \texttt{erb16}, candidate selection took 24\% of the time of \Edm{}, 135\,s, but the teardown and initialization of \Gab{} took 101 and 53\,s longer than those of \Edm{}, against 31 and 9.5\,s on the biclique blocks, which have a similar number of edges, 124 against 133 million. The bipartite constructions and the \texttt{erb} graphs have similar counts but differ in their numbering: in the constructions, the neighbours of a vertex have nearby numbers, whereas in an \texttt{erb} graph they are spread over the whole range. In an exploratory run on a smaller graph of biclique blocks, randomly relabelling the vertices reversed the ranking (Section~\ref{sec:threats}).

\emph{Expansion.} On the ER graph with the largest ratio, 1.19, \Gab{} spent 207\,s expanding blossoms, against 32\,s for \Edm{} on the same 51,358 expansions. In both programs an expansion splits the membership structure of the blossom and relabels its children. \Gab{} also adds the vertices of the children that become outer to the supporting forest and merges their blocks, and its expansions listed 11.3 billion vertex entries, against 1.1 million for \Edm{}. Contraction also cost \Gab{} more on this graph, 612 against 415\,s. On the other graphs, expansion took \Gab{} at most 3.6\,s longer than \Edm{}.

\emph{Contraction.} On the slow seeds, blossom contraction accounts for most of the extra time of \Edm{} and \Gab{}: 57--82\% of the difference for \Edm{} and 56--94\% for \Gab{}. On the slower 750,000-vertex graph, \Edm{} spent 1,828\,s of 3,074\,s contracting blossoms, against 3.6\,s on the faster seed. Each contraction lists the vertices of the new blossom (Section~\ref{sec:gabow-implementation}), and on this seed one blossom grew to 571,427 of the 750,000 vertices through nested contractions, up to a depth of 127,440; the contractions listed 47.7 billion vertex entries, counting a vertex again for every blossom it enters, 273 times as many as on the faster seed. In the 10 runs that listed more than a billion entries, contraction took 15--38\,ns per listed entry in \Edm{} and 26--52\,ns in \Gab{}, which also merges the blocks of the contracted blossoms. Contraction also separates the two seeds at 250,000 vertices (81\,s against 2.7\,s), the two at 1,024,000 vertices (415\,s against 9.6\,s), the two Watts--Strogatz seeds (36\,s against 0.6\,s) and the two weight ranges on one ER topology (119\,s with $W=10^3$ against 1.0\,s with $W=10^6$). \Scal{} took 277 and 288\,s on the two 750,000-vertex seeds and 86\,s on both Watts--Strogatz seeds, so the Watts--Strogatz reversal comes mainly from the contraction time of \Edm{}.

\emph{Weight scaling.} \Scal{} spent 75--98\% of its time in the Criterion-2 phases of FVR on every instrumented graph except the toroidal one; its Criterion-3 searches took at most 0.9\,s, dismantling at most 16\%, and each remaining stage, including Perfection, at most 4.5\% (Table~\ref{tab:profile-scaling}). On the toroidal graph, \Scal{} has two bottlenecks of different origin. Dismantling, including the preparation of the inherited blossoms, took 48\% of its 2,347\,s. At the start of a scale, before the large blossoms are liquidated, the inherited blossoms of the three toroidal graphs are nested up to 37,813--60,500 levels deep, against at most 359 on the other 47 Suite~B graphs. \Scal{} keeps, for every vertex, the list of inherited blossoms that contain it (Section~\ref{sec:scaling-implementation}), and it obtains the inherited duals of an edge by walking the part that the lists of the two endpoints share; on the profiled toroidal graph these walks took 1.1 trillion steps over all scales, about 4,850 per walk. The lists keep the liquidated blossoms, whose duals are zero, and the walks pass through them: blossoms with fewer than $\tau=13{,}452$ vertices can nest at most 6,725 levels deep, so at least 82\% of the entries of the longest list on each toroidal graph belong to large blossoms. We did not count how many of the walk steps pass through such entries. On the toroidal graphs the lists hold up to 7.1 billion entries, and their allocated capacity, 35.6--73.8\,GiB, nearly equals the peak memory of each graph, 36.9--71.0\,GiB. The internal timers of \Scal{}, which nest inside the stages of Table~\ref{tab:profile-scaling} and are not added to them, show where dismantling spent its time. Its 566,463 shell searches, against 241 on the ER graph of the same order and density, took 1,022\,s of the 1,044\,s of the dismantling stage. Of these, 876\,s went to the maximum-cardinality matchings that the shell searches compute: 289\,s to an initial greedy pass over tight edges, 234\,s to searches that found an augmenting path and 353\,s to final searches that found none, which ended 678,502 of the 1,003,668 matching computations. The Criterion-2 phases, which took the other 52\%, run after dismantling, when the lists have been released and no inherited blossom remains. Compared with the ER graph, they made a similar number of augmentations, 5.9 against 7.0 million over all scales, but scanned 10.9 against 1.9 billion edges, or 49 against 8.6 scans per edge and scale, and formed 1.65 million current blossoms against 120. Rebuilding the event state of a search after a blossom was expanded or dissolved took 253\,s on the toroidal graph and less than 0.1\,s on the ER graph; 98\% of these expansions and dissolutions occurred in the Criterion-2 phases, but the timer does not separate dismantling from FVR, so it does not show how the 253\,s divide between them.

\section{Discussion}
\label{sec:discussion}

\subsection{Interpreting the results}

Within one search, Gabow's candidate structure reduced running time only on graphs on which \Edm{} discards many stale queue entries per search step: on all 24 structured graphs, by up to 53\%, and on \texttt{spal\_004}, by 1.8\%. On the other 74 graphs it cost 1--39\% more time, and it needed 23--95\% more memory on every graph, so there the work of keeping its records current exceeded the saving in candidate selection. Many stale entries alone do not decide the outcome: \Gab{} remained slower on the random bipartite graphs of Suite~A, whose counts resemble those of the bipartite constructions. The phase times show the trade: the structure removes the cost of discarding stale candidates, which is small except on graphs such as \texttt{spal\_004} and the structured graphs, and adds the work of keeping its records current. Tree teardown, the largest extra cost on 40 of the 45 profiled graphs, belongs to our continuous search, which retires the records of finished trees instead of rebuilding the structure for each search as Gabow's design does (Section~\ref{sec:gabow-implementation}); we did not measure the per-search alternative. On the other five, the graphs whose contractions list more than a billion vertex entries, contraction was the largest extra cost, because \Gab{} also merges the blocks of the contracted blossoms. These results concern the implemented structure, including its binary-lifting NCA and its continuous search; they neither confirm nor refute the asymptotic improvement of Gabow's analysed design.

The ranking between \Scal{} and \Edm{} depends strongly on the graph family, and bipartiteness alone does not predict it: \Scal{} leads on bipartite ER graphs and \Edm{} on biclique blocks. The observations of Sections~\ref{sec:rq2} and~\ref{sec:profiles} give one account of the measured rankings: the classical programs are fast where their searches stay inside small components and slow where large blossoms are contracted repeatedly, and \Scal{} is slow on the geometric graph because of both its current blossoms and its representation of the inherited ones. The account describes the measured implementations, not general strengths or weaknesses of the published algorithms.

\subsection{Threats to validity}
\label{sec:threats}

\emph{Tie rule.} \Edm{} and \Gab{} perform the same search because they break ties by the same rule. Another rule can lead to another search and change the times of both programs. Ties are frequent here: with $W=1000$ and millions of edges, each weight value is shared by thousands of edges.

\emph{Run order.} The programs of a trial run back to back on the same node, in an order that alternates (Suite~A) or rotates (Suite~B) across trials, so no program always runs first.

\emph{Nodes.} All trials ran on nodes of the configuration of Section~\ref{sec:platform}, with the operating system's default memory placement. On three of the 23 nodes, runs whose peak memory exceeded about 60\,GiB took up to 1.9 times as long as the same runs on other nodes, which ran processes of up to 137\,GiB without such delays: on \texttt{erb18}, \Edm{} took 1,352\,s and \Gab{} 1,879\,s on such a node, against 909 and 987\,s on another. All 15 runs that took more than 10\% longer than the median of their graph and program ran on these three nodes, on large Suite~A graphs. Both programs of a trial run on the same node, but \Gab{} needs more memory, so these nodes raised the per-trial ratio without reversing it: \Edm{} was faster in every trial of every SuiteSparse and generated graph of Suite~A except \texttt{spal\_004}. Over the 24 of these graphs that also ran on other nodes, the geometric mean of the ratio is 1.162 with all trials and 1.130 without the trials on the three nodes, and the ratios of the random bipartite graphs fall from 1.01--1.39 to 1.01--1.12. All trials of \texttt{erb8} ran on one of the three nodes, so its ratio, 1.24, may be raised by this effect. The memory configuration of the nodes was not recorded, so we cannot attribute the effect further.

\emph{Numbering.} In the structured graphs, the neighbours of a vertex have nearby numbers (Section~\ref{sec:suite-a}), whereas in a random graph they are spread over all numbers. The programs store the data of a vertex at a position given by its number, so nearby numbers keep the data of adjacent vertices close together in memory, where the processor's caches serve it faster. We did not measure the structured graphs with randomly permuted vertex numbers. In an exploratory single run on a smaller graph of biclique blocks, with 67,584 vertices, relabelling the vertices of each side at random made \Gab{} 47\% slower than \Edm{}, whereas it had been 12\% faster. Relabelling also changes the order in which ties are broken, and with it the search, so that run does not isolate the effect of the numbering; but the advantage of \Gab{} on the bipartite constructions may depend on it.

\emph{Selection.} The graphs were fixed after earlier runs, and graphs on which \Edm{} or \Gab{} did not finish were left out; they include ER graphs with $W=10$, which \Scal{} solved. The structured graphs were built to look for graphs on which \Gab{} is faster, so their results show that such graphs exist, not how common they are.

\emph{Scope.} Our conclusions concern the selected graphs and their stored weights. Weight magnitude and tie frequency vary together, and the block graphs change several structural properties at once. Trial ranges describe repeatability on one graph and seed ranges describe variation between instances; neither is a confidence statement about other graphs. The three programs are research implementations built to isolate candidate management and weight scaling; we do not compare them with production libraries, and LEMON serves only as a correctness oracle. Phase times cover the 21 instrumented graphs and the 24 structured graphs; operation counts cover the 21 graphs, all 50 Suite~B graphs for \Scal{}, and the search steps and queue entries of \Edm{} and \Gab{} on all 99 graphs (Section~\ref{sec:validation}). The explanations of Section~\ref{sec:profiles} rest mainly on these 45 profiled graphs.

\section{Related Work}
\label{sec:related}

Edmonds introduced blossom contraction and the polyhedral description of matchings on which weighted general matching rests~\cite{EdmondsWeighted1965,EdmondsFlowers1965}. Micali and Vazirani obtained $O(m\sqrt n)$ time for maximum-cardinality matching~\cite{MicaliVazirani1980}; this is background, not one of the weighted problems measured here.

For weighted matching, Gabow introduced a scaling algorithm in 1985~\cite{Gabow1985}. Galil, Micali and Gabow gave the $O(mn\log n)$ implementation with generalized priority queues and concatenable queues on which \Edm{} is modelled~\cite{GalilMicaliGabow1986}, and Gabow, Galil and Spencer improved the bound with data structures based on contraction~\cite{GabowGalilSpencer1989}. Gabow's supporting-forest data structures, introduced in 1990 and developed for weighted matching, $b$-matching and $f$-factors, give $O(n(m+n\log n))$~\cite{Gabow1990,Gabow2018}; they rely on his structure for nearest common ancestors with linking~\cite{Gabow2017}. \Gab{} is modelled on this line of work. A public $O(nm\log n)$ solver by Min\_25, which serves as the reference solution of the general weighted matching problem of Library Checker, cites Gabow's paper but keeps the edges between outer vertices in one global priority queue, as \Edm{} does~\cite{Min25}.

Gabow and Tarjan's scaling algorithm~\cite{GabowTarjan1991} was the direct predecessor of the algorithm of Duan, Pettie and Su~\cite{DuanPettieSu2018}. Their Hybrid algorithm combines the liquidation of large blossoms, a method derived from Gabow's 1985 algorithm for dismantling small inherited blossoms, free-vertex reduction and bounded searches. Duan, Pettie and Su call that dismantling method Gabow's algorithm; it is distinct from the program we call \Gab{}.

Cook and Rohe's Blossom~IV, a practical minimum-weight perfect-matching solver, adjusts the duals of each search tree separately~\cite{CookRohe1999}. Mehlhorn and Sch\"afer implement an $O(nm\log n)$ weighted algorithm and study whether sophisticated data structures improve actual running time, comparing against Blossom~IV~\cite{MehlhornSchaefer2002}. Kolmogorov's Blossom~V combines these per-tree dual adjustments with priority queues and compares against both~\cite{Kolmogorov2009}, and Arkhipov and Kolmogorov's Blossom~VI uses cherry-tree searches and delayed blossom contraction~\cite{ArkhipovKolmogorov2026}. These studies are the closest precedents for relating implementation choices to measured behaviour.

Recent studies of cardinality matching ask a related engineering question. Ansaripour, Danaei and Mehlhorn find that Gabow's asymptotic improvement for cardinality matching pays off on their worst-case families but not on random graphs~\cite{AnsaripourDanaeiMehlhorn2024}, and Mehlhorn and Nobahari revisit Gabow's $O(\sqrt n\,m)$ cardinality algorithm with a new way of computing the lengths of shortest augmenting paths~\cite{MehlhornNobahari2026}. Both concern maximum cardinality, not the weighted supporting-forest method or Hybrid scaling. LEMON provides the maximum-weight and maximum-weight perfect matching solvers that we use as correctness oracles~\cite{LEMON2011}.

The gap we address is the practical cost of Gabow's supporting-forest candidate management relative to the organization of Galil, Micali and Gabow, and of both relative to the scaling algorithm of Duan, Pettie and Su. Mehlhorn and Sch\"afer study data structures within the $O(nm\log n)$ family, and Blossom~IV, V and VI study other organizations of perfect-matching search. Our design pairs \Edm{} and \Gab{} within one search and runs all three programs on the same stored graphs whenever an independent solver shows that their optima coincide.

\section{Conclusion}
\label{sec:conclusion}

We compared three exact programs for maximum weighted matching on general graphs. \Edm{} and \Gab{} performed the same search on every graph, and within it Gabow's supporting-forest candidate structure paid off on the structured graphs of Suite~A, by up to 53\%, and on one SuiteSparse graph, by 1.8\%, whereas the single lazily pruned queue of \Edm{} was faster on the other 74 graphs; \Gab{} needed 23--95\% more memory throughout. The Hybrid scaling program was 2.6--11.8 times faster than \Edm{} on Erd\H{o}s--R\'enyi graphs and also led on bipartite random and community graphs, but it was 26--33 times slower on random geometric graphs, where it also needed up to 22 times the memory, and slower on disconnected clique and biclique blocks. For these implementations and graphs, the family mattered more than the asymptotic bounds: \Scal{} was faster on the random sparse graphs, and \Edm{} on the geometric and block-structured ones. Instrumented runs locate these differences in the work of each program: \Gab{} gains where \Edm{} discards many stale queue entries per search step, although many stale entries alone did not make it faster, and its largest extra cost was tree teardown, or contraction on random graphs with deeply nested blossoms; repeated contraction of large blossoms makes \Edm{} and \Gab{} sensitive to the seed; and on geometric graphs \Scal{} forms far more blossoms in its searches and spends about half of its time dismantling deeply nested inherited blossoms.

The practical solvers cited here do not start the blossom search from an empty matching. Cook and Rohe initialize the matching and the duals from a linear-programming relaxation~\cite{CookRohe1999}, and Blossom~VI starts from feasible duals and a matching on tight edges obtained by a fractional initialization~\cite{ArkhipovKolmogorov2026}. We kept the simple per-vertex start so that measured differences come from the candidate structures and the search itself; combining these structures with such a warm start is a further step towards a practical code.

\bibliographystyle{ACM-Reference-Format}
\bibliography{references}


\begin{thebibliography}{28}


\ifx \showCODEN    \undefined \def \showCODEN     #1{\unskip}     \fi
\ifx \showISBNx    \undefined \def \showISBNx     #1{\unskip}     \fi
\ifx \showISBNxiii \undefined \def \showISBNxiii  #1{\unskip}     \fi
\ifx \showISSN     \undefined \def \showISSN      #1{\unskip}     \fi
\ifx \showLCCN     \undefined \def \showLCCN      #1{\unskip}     \fi
\ifx \shownote     \undefined \def \shownote      #1{#1}          \fi
\ifx \showarticletitle \undefined \def \showarticletitle #1{#1}   \fi
\ifx \showURL      \undefined \def \showURL       {\relax}        \fi
\providecommand\bibfield[2]{#2}
\providecommand\bibinfo[2]{#2}
\providecommand\natexlab[1]{#1}
\providecommand\showeprint[2][]{arXiv:#2}
\makeatletter
\@ifundefined{NAT@parse@date}{}{\let\NAT@parse@date@orig\NAT@parse@date}
\@ifundefined{NAT@parse@date}{}{\def\NAT@parse@date#1#2#3#4#5#6@@{\NAT@parse@date@orig#1#2#3#4#5#6@@\def\NAT@tempyear{0000}\def\NAT@tempexlab{{?}}\ifx\NAT@year\NAT@tempyear\ifx\NAT@exlab\NAT@tempexlab\def\NAT@date{[n.\,d.]}\else\edef\NAT@date{[n.\,d.]\NAT@exlab}\fi\fi}}
\makeatother

\bibitem[Ansaripour et~al\mbox{.}(2024)]%
        {AnsaripourDanaeiMehlhorn2024}
\bibfield{author}{\bibinfo{person}{Matin Ansaripour}, \bibinfo{person}{Alireza
  Danaei}, {and} \bibinfo{person}{Kurt Mehlhorn}.}
  \bibinfo{year}{2024}\natexlab{}.
\newblock \bibinfo{title}{Gabow's Cardinality Matching Algorithm in General
  Graphs: Implementation and Experiments}.
\newblock
\showeprint[arxiv]{2409.14849}~[cs.DS]


\bibitem[Arkhipov and Kolmogorov(2026)]%
        {ArkhipovKolmogorov2026}
\bibfield{author}{\bibinfo{person}{Pavel Arkhipov} {and}
  \bibinfo{person}{Vladimir Kolmogorov}.} \bibinfo{year}{2026}\natexlab{}.
\newblock \bibinfo{title}{Blossom {VI}: A Practical Minimum Weight Perfect
  Matching Algorithm}.
\newblock
\showeprint[arxiv]{2604.20351}~[cs.DS]
\newblock
\shownote{Version 2, revised 21 July 2026. Implementation:
  \url{https://github.com/Paul566/blossom-vi}}.


\bibitem[Bader et~al\mbox{.}(2013)]%
        {Bader2013}
\bibfield{editor}{\bibinfo{person}{David~A. Bader}, \bibinfo{person}{Henning
  Meyerhenke}, \bibinfo{person}{Peter Sanders}, {and} \bibinfo{person}{Dorothea
  Wagner}} (Eds.). \bibinfo{year}{2013}\natexlab{}.
\newblock \bibinfo{booktitle}{\emph{Graph Partitioning and Graph Clustering:
  10th {DIMACS} Implementation Challenge Workshop}}.
  \bibinfo{series}{Contemporary Mathematics}, Vol.~\bibinfo{volume}{588}.
\newblock \bibinfo{publisher}{American Mathematical Society},
  \bibinfo{address}{Providence, RI}.
\newblock
\href{https://doi.org/10.1090/conm/588}{doi:\nolinkurl{10.1090/conm/588}}


\bibitem[Barab{\'a}si and Albert(1999)]%
        {BarabasiAlbert1999}
\bibfield{author}{\bibinfo{person}{Albert-L{\'a}szl{\'o} Barab{\'a}si} {and}
  \bibinfo{person}{R{\'e}ka Albert}.} \bibinfo{year}{1999}\natexlab{}.
\newblock \showarticletitle{Emergence of Scaling in Random Networks}.
\newblock \bibinfo{journal}{\emph{Science}} \bibinfo{volume}{286},
  \bibinfo{number}{5439} (\bibinfo{year}{1999}), \bibinfo{pages}{509--512}.
\newblock
\href{https://doi.org/10.1126/science.286.5439.509}{doi:\nolinkurl{10.1126/science.286.5439.509}}


\bibitem[Cook and Rohe(1999)]%
        {CookRohe1999}
\bibfield{author}{\bibinfo{person}{William Cook} {and}
  \bibinfo{person}{Andr{\'e} Rohe}.} \bibinfo{year}{1999}\natexlab{}.
\newblock \showarticletitle{Computing Minimum-Weight Perfect Matchings}.
\newblock \bibinfo{journal}{\emph{INFORMS Journal on Computing}}
  \bibinfo{volume}{11}, \bibinfo{number}{2} (\bibinfo{year}{1999}),
  \bibinfo{pages}{138--148}.
\newblock
\href{https://doi.org/10.1287/ijoc.11.2.138}{doi:\nolinkurl{10.1287/ijoc.11.2.138}}


\bibitem[Davis and Hu(2011)]%
        {DavisHu2011}
\bibfield{author}{\bibinfo{person}{Timothy~A. Davis} {and}
  \bibinfo{person}{Yifan Hu}.} \bibinfo{year}{2011}\natexlab{}.
\newblock \showarticletitle{The {U}niversity of {F}lorida Sparse Matrix
  Collection}.
\newblock \bibinfo{journal}{\emph{ACM Trans. Math. Software}}
  \bibinfo{volume}{38}, \bibinfo{number}{1}, Article \bibinfo{articleno}{1}
  (\bibinfo{year}{2011}), \bibinfo{numpages}{25}~pages.
\newblock
\href{https://doi.org/10.1145/2049662.2049663}{doi:\nolinkurl{10.1145/2049662.2049663}}


\bibitem[Dezs{\H{o}} et~al\mbox{.}(2011)]%
        {LEMON2011}
\bibfield{author}{\bibinfo{person}{Bal{\'a}zs Dezs{\H{o}}},
  \bibinfo{person}{Alp{\'a}r J{\"u}ttner}, {and} \bibinfo{person}{P{\'e}ter
  Kov{\'a}cs}.} \bibinfo{year}{2011}\natexlab{}.
\newblock \showarticletitle{{LEMON}---An Open Source {C++} Graph Template
  Library}.
\newblock \bibinfo{journal}{\emph{Electronic Notes in Theoretical Computer
  Science}} \bibinfo{volume}{264}, \bibinfo{number}{5} (\bibinfo{year}{2011}),
  \bibinfo{pages}{23--45}.
\newblock
\href{https://doi.org/10.1016/j.entcs.2011.06.003}{doi:\nolinkurl{10.1016/j.entcs.2011.06.003}}


\bibitem[Duan et~al\mbox{.}(2018)]%
        {DuanPettieSu2018}
\bibfield{author}{\bibinfo{person}{Ran Duan}, \bibinfo{person}{Seth Pettie},
  {and} \bibinfo{person}{Hsin-Hao Su}.} \bibinfo{year}{2018}\natexlab{}.
\newblock \showarticletitle{Scaling Algorithms for Weighted Matching in General
  Graphs}.
\newblock \bibinfo{journal}{\emph{ACM Transactions on Algorithms}}
  \bibinfo{volume}{14}, \bibinfo{number}{1}, Article \bibinfo{articleno}{8}
  (\bibinfo{year}{2018}), \bibinfo{numpages}{35}~pages.
\newblock
\href{https://doi.org/10.1145/3155301}{doi:\nolinkurl{10.1145/3155301}}
\newblock
\shownote{Also available as arXiv:1411.1919, whose numbering of properties and
  lemmas we follow}.


\bibitem[Edmonds(1965a)]%
        {EdmondsWeighted1965}
\bibfield{author}{\bibinfo{person}{Jack Edmonds}.}
  \bibinfo{year}{1965}\natexlab{a}.
\newblock \showarticletitle{Maximum Matching and a Polyhedron with
  0,1-Vertices}.
\newblock \bibinfo{journal}{\emph{Journal of Research of the National Bureau of
  Standards, Section B: Mathematics and Mathematical Physics}}
  \bibinfo{volume}{69B}, \bibinfo{number}{1--2} (\bibinfo{year}{1965}),
  \bibinfo{pages}{125--130}.
\newblock
\href{https://doi.org/10.6028/jres.069B.013}{doi:\nolinkurl{10.6028/jres.069B.013}}


\bibitem[Edmonds(1965b)]%
        {EdmondsFlowers1965}
\bibfield{author}{\bibinfo{person}{Jack Edmonds}.}
  \bibinfo{year}{1965}\natexlab{b}.
\newblock \showarticletitle{Paths, Trees, and Flowers}.
\newblock \bibinfo{journal}{\emph{Canadian Journal of Mathematics}}
  \bibinfo{volume}{17} (\bibinfo{year}{1965}), \bibinfo{pages}{449--467}.
\newblock
\href{https://doi.org/10.4153/CJM-1965-045-4}{doi:\nolinkurl{10.4153/CJM-1965-045-4}}


\bibitem[Erd{\H{o}}s and R{\'e}nyi(1959)]%
        {ErdosRenyi1959}
\bibfield{author}{\bibinfo{person}{Paul Erd{\H{o}}s} {and}
  \bibinfo{person}{Alfr{\'e}d R{\'e}nyi}.} \bibinfo{year}{1959}\natexlab{}.
\newblock \showarticletitle{On Random Graphs {I}}.
\newblock \bibinfo{journal}{\emph{Publicationes Mathematicae Debrecen}}
  \bibinfo{volume}{6} (\bibinfo{year}{1959}), \bibinfo{pages}{290--297}.
\newblock
\href{https://doi.org/10.5486/PMD.1959.6.3-4.12}{doi:\nolinkurl{10.5486/PMD.1959.6.3-4.12}}


\bibitem[Gabow(1985)]%
        {Gabow1985}
\bibfield{author}{\bibinfo{person}{Harold~N. Gabow}.}
  \bibinfo{year}{1985}\natexlab{}.
\newblock \showarticletitle{A Scaling Algorithm for Weighted Matching on
  General Graphs}. In \bibinfo{booktitle}{\emph{Proceedings of the 26th Annual
  Symposium on Foundations of Computer Science}}. \bibinfo{publisher}{IEEE
  Computer Society}, \bibinfo{address}{Los Alamitos, CA},
  \bibinfo{pages}{90--100}.
\newblock
\href{https://doi.org/10.1109/SFCS.1985.3}{doi:\nolinkurl{10.1109/SFCS.1985.3}}


\bibitem[Gabow(1990)]%
        {Gabow1990}
\bibfield{author}{\bibinfo{person}{Harold~N. Gabow}.}
  \bibinfo{year}{1990}\natexlab{}.
\newblock \showarticletitle{Data Structures for Weighted Matching and Nearest
  Common Ancestors with Linking}. In \bibinfo{booktitle}{\emph{Proceedings of
  the First Annual ACM--SIAM Symposium on Discrete Algorithms}}.
  \bibinfo{publisher}{SIAM}, \bibinfo{address}{Philadelphia, PA},
  \bibinfo{pages}{434--443}.
\newblock


\bibitem[Gabow(2017)]%
        {Gabow2017}
\bibfield{author}{\bibinfo{person}{Harold~N. Gabow}.}
  \bibinfo{year}{2017}\natexlab{}.
\newblock \showarticletitle{A Data Structure for Nearest Common Ancestors with
  Linking}.
\newblock \bibinfo{journal}{\emph{ACM Transactions on Algorithms}}
  \bibinfo{volume}{13}, \bibinfo{number}{4}, Article \bibinfo{articleno}{45}
  (\bibinfo{year}{2017}), \bibinfo{numpages}{28}~pages.
\newblock
\href{https://doi.org/10.1145/3108240}{doi:\nolinkurl{10.1145/3108240}}


\bibitem[Gabow(2018)]%
        {Gabow2018}
\bibfield{author}{\bibinfo{person}{Harold~N. Gabow}.}
  \bibinfo{year}{2018}\natexlab{}.
\newblock \showarticletitle{Data Structures for Weighted Matching and
  Extensions to $b$-Matching and $f$-Factors}.
\newblock \bibinfo{journal}{\emph{ACM Transactions on Algorithms}}
  \bibinfo{volume}{14}, \bibinfo{number}{3}, Article \bibinfo{articleno}{39}
  (\bibinfo{year}{2018}), \bibinfo{numpages}{80}~pages.
\newblock
\href{https://doi.org/10.1145/3183369}{doi:\nolinkurl{10.1145/3183369}}


\bibitem[Gabow et~al\mbox{.}(1989)]%
        {GabowGalilSpencer1989}
\bibfield{author}{\bibinfo{person}{Harold~N. Gabow}, \bibinfo{person}{Zvi
  Galil}, {and} \bibinfo{person}{Thomas~H. Spencer}.}
  \bibinfo{year}{1989}\natexlab{}.
\newblock \showarticletitle{Efficient Implementation of Graph Algorithms Using
  Contraction}.
\newblock \bibinfo{journal}{\emph{J. ACM}} \bibinfo{volume}{36},
  \bibinfo{number}{3} (\bibinfo{year}{1989}), \bibinfo{pages}{540--572}.
\newblock
\href{https://doi.org/10.1145/65950.65954}{doi:\nolinkurl{10.1145/65950.65954}}


\bibitem[Gabow and Tarjan(1991)]%
        {GabowTarjan1991}
\bibfield{author}{\bibinfo{person}{Harold~N. Gabow} {and}
  \bibinfo{person}{Robert~E. Tarjan}.} \bibinfo{year}{1991}\natexlab{}.
\newblock \showarticletitle{Faster Scaling Algorithms for General Graph
  Matching Problems}.
\newblock \bibinfo{journal}{\emph{J. ACM}} \bibinfo{volume}{38},
  \bibinfo{number}{4} (\bibinfo{year}{1991}), \bibinfo{pages}{815--853}.
\newblock
\href{https://doi.org/10.1145/115234.115366}{doi:\nolinkurl{10.1145/115234.115366}}


\bibitem[Galil(1986)]%
        {Galil1986Survey}
\bibfield{author}{\bibinfo{person}{Zvi Galil}.}
  \bibinfo{year}{1986}\natexlab{}.
\newblock \showarticletitle{Efficient Algorithms for Finding Maximum Matching
  in Graphs}.
\newblock \bibinfo{journal}{\emph{Comput. Surveys}} \bibinfo{volume}{18},
  \bibinfo{number}{1} (\bibinfo{year}{1986}), \bibinfo{pages}{23--38}.
\newblock
\href{https://doi.org/10.1145/6462.6502}{doi:\nolinkurl{10.1145/6462.6502}}


\bibitem[Galil et~al\mbox{.}(1986)]%
        {GalilMicaliGabow1986}
\bibfield{author}{\bibinfo{person}{Zvi Galil}, \bibinfo{person}{Silvio Micali},
  {and} \bibinfo{person}{Harold~N. Gabow}.} \bibinfo{year}{1986}\natexlab{}.
\newblock \showarticletitle{An ${O}({EV}\log {V})$ Algorithm for Finding a
  Maximal Weighted Matching in General Graphs}.
\newblock \bibinfo{journal}{\emph{SIAM J. Comput.}} \bibinfo{volume}{15},
  \bibinfo{number}{1} (\bibinfo{year}{1986}), \bibinfo{pages}{120--130}.
\newblock
\href{https://doi.org/10.1137/0215009}{doi:\nolinkurl{10.1137/0215009}}


\bibitem[Gilbert(1959)]%
        {Gilbert1959}
\bibfield{author}{\bibinfo{person}{Edgar~N. Gilbert}.}
  \bibinfo{year}{1959}\natexlab{}.
\newblock \showarticletitle{Random Graphs}.
\newblock \bibinfo{journal}{\emph{The Annals of Mathematical Statistics}}
  \bibinfo{volume}{30}, \bibinfo{number}{4} (\bibinfo{year}{1959}),
  \bibinfo{pages}{1141--1144}.
\newblock
\href{https://doi.org/10.1214/aoms/1177706098}{doi:\nolinkurl{10.1214/aoms/1177706098}}


\bibitem[Holland et~al\mbox{.}(1983)]%
        {HollandLaskeyLeinhardt1983}
\bibfield{author}{\bibinfo{person}{Paul~W. Holland},
  \bibinfo{person}{Kathryn~Blackmond Laskey}, {and} \bibinfo{person}{Samuel
  Leinhardt}.} \bibinfo{year}{1983}\natexlab{}.
\newblock \showarticletitle{Stochastic Blockmodels: First Steps}.
\newblock \bibinfo{journal}{\emph{Social Networks}} \bibinfo{volume}{5},
  \bibinfo{number}{2} (\bibinfo{year}{1983}), \bibinfo{pages}{109--137}.
\newblock
\href{https://doi.org/10.1016/0378-8733(83)90021-7}{doi:\nolinkurl{10.1016/0378-8733(83)90021-7}}


\bibitem[Kolmogorov(2009)]%
        {Kolmogorov2009}
\bibfield{author}{\bibinfo{person}{Vladimir Kolmogorov}.}
  \bibinfo{year}{2009}\natexlab{}.
\newblock \showarticletitle{Blossom {V}: A New Implementation of a Minimum Cost
  Perfect Matching Algorithm}.
\newblock \bibinfo{journal}{\emph{Mathematical Programming Computation}}
  \bibinfo{volume}{1}, \bibinfo{number}{1} (\bibinfo{year}{2009}),
  \bibinfo{pages}{43--67}.
\newblock
\href{https://doi.org/10.1007/s12532-009-0002-8}{doi:\nolinkurl{10.1007/s12532-009-0002-8}}


\bibitem[Leskovec and Krevl(2014)]%
        {LeskovecKrevl2014}
\bibfield{author}{\bibinfo{person}{Jure Leskovec} {and} \bibinfo{person}{Andrej
  Krevl}.} \bibinfo{year}{2014}\natexlab{}.
\newblock \bibinfo{title}{{SNAP} Datasets: {S}tanford Large Network Dataset
  Collection}.
\newblock \bibinfo{howpublished}{\url{http://snap.stanford.edu/data}}.
\newblock


\bibitem[Mehlhorn and Nobahari(2026)]%
        {MehlhornNobahari2026}
\bibfield{author}{\bibinfo{person}{Kurt Mehlhorn} {and} \bibinfo{person}{Romina
  Nobahari}.} \bibinfo{year}{2026}\natexlab{}.
\newblock \bibinfo{title}{Gabow's ${O}(\sqrt{n}m)$ Maximum Cardinality Matching
  Algorithm, Revisited}.
\newblock
\showeprint[arxiv]{2603.22909}~[cs.DS]
\newblock
\shownote{Version 2, revised 21 April 2026}.


\bibitem[Mehlhorn and Sch{\"a}fer(2002)]%
        {MehlhornSchaefer2002}
\bibfield{author}{\bibinfo{person}{Kurt Mehlhorn} {and} \bibinfo{person}{Guido
  Sch{\"a}fer}.} \bibinfo{year}{2002}\natexlab{}.
\newblock \showarticletitle{Implementation of ${O}(nm\log n)$ Weighted
  Matchings in General Graphs: The Power of Data Structures}.
\newblock \bibinfo{journal}{\emph{ACM Journal of Experimental Algorithmics}}
  \bibinfo{volume}{7} (\bibinfo{year}{2002}), \bibinfo{pages}{4}.
\newblock
\href{https://doi.org/10.1145/944618.944622}{doi:\nolinkurl{10.1145/944618.944622}}


\bibitem[Micali and Vazirani(1980)]%
        {MicaliVazirani1980}
\bibfield{author}{\bibinfo{person}{Silvio Micali} {and}
  \bibinfo{person}{Vijay~V. Vazirani}.} \bibinfo{year}{1980}\natexlab{}.
\newblock \showarticletitle{An ${O}(\sqrt{|{V}|}\,|{E}|)$ Algorithm for Finding
  Maximum Matching in General Graphs}. In \bibinfo{booktitle}{\emph{Proceedings
  of the 21st Annual Symposium on Foundations of Computer Science}}.
  \bibinfo{publisher}{IEEE Computer Society}, \bibinfo{address}{Los Alamitos,
  CA}, \bibinfo{pages}{17--27}.
\newblock
\href{https://doi.org/10.1109/SFCS.1980.12}{doi:\nolinkurl{10.1109/SFCS.1980.12}}


\bibitem[{Min\_25}(2021)]%
        {Min25}
\bibfield{author}{\bibinfo{person}{{Min\_25}}.}
  \bibinfo{year}{2021}\natexlab{}.
\newblock \bibinfo{title}{Maximum Weighted Matching in General Graphs}.
\newblock
  \bibinfo{howpublished}{\url{https://github.com/yosupo06/library-checker-problems/blob/master/graph/general_weighted_matching/sol/correct.cpp}}.
\newblock
\newblock
\shownote{Source code; reference solution of the general weighted matching
  problem of Library Checker. Accessed 5 November 2026}.


\bibitem[Watts and Strogatz(1998)]%
        {WattsStrogatz1998}
\bibfield{author}{\bibinfo{person}{Duncan~J. Watts} {and}
  \bibinfo{person}{Steven~H. Strogatz}.} \bibinfo{year}{1998}\natexlab{}.
\newblock \showarticletitle{Collective Dynamics of `Small-World' Networks}.
\newblock \bibinfo{journal}{\emph{Nature}}  \bibinfo{volume}{393}
  (\bibinfo{year}{1998}), \bibinfo{pages}{440--442}.
\newblock
\href{https://doi.org/10.1038/30918}{doi:\nolinkurl{10.1038/30918}}


\end{thebibliography}

\appendix
\section{Detailed Scaling Specification}
\label{app:scaling-subroutines}

This appendix states the conditions that \Scal{} maintains, the Perfection argument, the shell notation and the control flow of dismantling.

\subsection{Relaxed complementary slackness and exactness}
\label{app:scaling-invariants}

For a blossom $B$, $E_B$ is its edge set (Section~\ref{sec:blossoms}) and $E(B)$ the set of all input edges inside $B$. The two sets differ in general: for a blossom formed by a 5-cycle with a chord, $E_B$ holds the five cycle edges and $E(B)$ also the chord. Once the inherited blossoms are gone, the current state satisfies relaxed complementary slackness, Property~2 of Duan, Pettie and Su; while inherited and current blossoms coexist, their Property~3 governs the interaction, and an inherited blossom may still have positive dual~\cite{DuanPettieSu2018}. Vertex duals are integers and blossom duals are nonnegative even integers. Every $B\in\Omega$ is full, $|M\cap E_B|=\lfloor|B|/2\rfloor$, which implies $M\cap E(B)=M\cap E_B$, since no matching has more than $\lfloor|B|/2\rfloor$ edges inside $B$. Every maximal blossom of $\Omega$ has positive dual, a nonmaximal blossom may have zero dual, and $z(B)=0$ for every $B\in\mathcal O\setminus\Omega$. In addition,
\begin{align}
  yz(e)&\geq w(e)-2 &&\text{for every edge }e \quad\text{(near domination)},\\
  yz(e)&\leq w(e) &&\text{for every }e\in M\cup\bigcup_{B\in\Omega}E_B
       \quad\text{(near tightness)}.
\end{align}
The relaxation unit is two because the weights of every scale are even. If these conditions hold for a perfect matching, its weight is within $n$ of the optimum for the weights of the scale. In the final scale $w=2\bar w$, so the matching is within $n/2$ of the optimum under $\bar w$. Matching weights under $\bar w$ are multiples of $n/2+1>n/2$, so the matching is optimal under $\bar w$ and hence under $\widehat w$~\cite{DuanPettieSu2018}.

\subsection{Perfection}
\label{app:perfection}

Consider the state after the free dummies of earlier scales have been deleted and before new dummies are added. Assume that this state satisfies the conditions above, that every remaining dummy is matched, and that each new dummy lies outside every current blossom. For each free real vertex $u$, add a new vertex $\hat u$, the edge $u\hat u$ of weight zero and the dual $y(\hat u)=-y(u)$, and match the pair. All old duals, blossoms and matching edges are left unchanged.

This operation preserves relaxed complementary slackness. Every old edge has the same endpoints, containing blossoms and dual value as before, so its domination and tightness conditions are unchanged. The new edge lies in no blossom and has
\begin{equation}
  yz(u\hat u)=y(u)+y(\hat u)=0=w(u\hat u),
\end{equation}
so it is both dominated and tight. Each new matching edge joins two previously free vertices, so the matching remains valid and now covers every vertex. No new edge enters an old edge set $E_B$, so every blossom stays full. The laminar family and its duals are unchanged, which preserves the nonnegativity and evenness of blossom duals and the positivity of maximal ones. The new vertex dual is an integer because $y(u)$ is one, and \MWPM{} imposes no sign restriction on vertex duals. If the free real vertices have duals of a common parity, the new dummy duals have the same parity, because $-y(u)\equiv y(u)\pmod 2$. The argument is local: it assumes that the state before the insertion satisfies the stated conditions.

\subsection{Major paths, shells and eligibility}
\label{app:scaling-shells}

The inherited forest is decomposed into major paths, as in Figure~\ref{fig:major-path-shell}. A child $B_1$ of $B_2$ is major when $|B_1|>|B_2|/2$; a node $R$ that is not a major child starts the path $P(R)$ obtained by repeatedly moving to the unique major child. Major-path roots are processed in postorder. For undissolved inherited boundaries $D\subset C$ on $P(R)$, the shell $G(C,D)$ is the subgraph induced by $C\setminus D$, and it is atomic when no undissolved inherited boundary lies strictly between $D$ and $C$. During \textsc{ShellSearch}, $C^*$ is the smallest undissolved inherited blossom containing $C$, $D^*$ is the largest undissolved inherited blossom contained in $D$ (or $\varnothing$), and $G(C^*,D^*)$ is the current shell. Within \textsc{DismantlePath}, $F$ denotes the free vertices still inside undissolved blossoms of $P(R)$, and $F^*$ the free vertices of $G(C^*,D^*)$.

Each unit shell translation applies
\begin{equation}
\begin{aligned}
z(C^*)&\leftarrow z(C^*)-2, & z(D^*)&\leftarrow z(D^*)-2 &&(D^*\neq\varnothing),\\
y(u)&\leftarrow y(u)+2 &&& &(u\in D^*),\\
y(u)&\leftarrow y(u)+1 &&& &(u\in C^*\setminus D^*).
\end{aligned}
\end{equation}
In terms of the slack of Equation~\eqref{eq:edge-slack}, Criterion~1 admits tight edges, $\pi(e)=0$. Criterion~2 admits the edges $\bigcup_{B\in\Omega}E_B$ of current blossoms, unmatched edges with $\pi(e)=-2$ and matched edges with $\pi(e)=0$. Criterion~3 admits every edge with $\pi(e)\in\{0,-2\}$. Under Criterion~2, an edge between blossoms that has just been matched has $yz(e)=w(e)-2$: it satisfies near tightness but is no longer eligible. Augmenting along a path therefore makes all of its edges ineligible, so a maximal family of vertex-disjoint augmenting paths removes every eligible augmenting path from the search roots (Lemma~2.3 of~\cite{DuanPettieSu2018}). Criterion~3 keeps the former edges of a current blossom eligible when that blossom is expanded during a search.

\subsection{Dismantling control flow}
\label{app:scaling-control}

\textsc{DismantlePath} removes the inherited blossoms along one major path. Let $\mathcal L=P(R)\cap\Omega'$ be its chain of undissolved inherited blossoms. For $B_1\subset\cdots\subset B_k$ in $\mathcal L$, with $B_0=\varnothing$, the atomic shells are $G(B_j,B_{j-1})$. Each iteration searches the shells with at least two free vertices, in decreasing order of that number. The shared set $A$ records the vertices searched by earlier calls in the same iteration.

If an undissolved blossom remains when this loop stops, it contains exactly one free vertex. Indeed, the largest undissolved boundary has odd size and no matched edge crosses an inherited boundary, so it contains an odd number of free vertices; the loop stops when fewer than two remain, which leaves a unique vertex $\omega$. The same parity argument places $\omega$ in every remaining undissolved blossom. Liquidating those blossoms increases $y(\omega)$ by $h=\tfrac12\sum_{B\in\mathcal L}z(B)$, and a Criterion-1 search from $\omega$ for exactly $h$ dual units restores that increase; under the dismantling conditions it finds no augmenting path~\cite{DuanPettieSu2018}. This final search grows and contracts along tight reachable edges, adjusts duals and expands enabled current blossoms, and stops at its dual budget; it is the bounded use of $\operatorname{PQSearch}$ by Duan, Pettie and Su.

\begin{figure}[tbp]
\begin{pseudocode}{Algorithm 4}{DismantlePath$(R)$}
  \pseudometa{Input}{A major-path root $R$ whose off-path descendants are already dismantled.}
  \pseudometa{Output}{All inherited blossoms on $P(R)$ are dissolved.}
  \pline{$\mathcal L\leftarrow P(R)\cap\Omega'$.}
  \pline{$F\leftarrow$ the free vertices in $\bigcup\mathcal L$.}
  \pline{\textbf{while} $\mathcal L\neq\varnothing$ and $|F|\geq2$ \textbf{do}}
  \pline[1]{Form the atomic shells $G(C,D)$ between consecutive members of $\mathcal L$.}
  \pline[1]{Keep the shells with at least two free vertices; sort them by decreasing number of free vertices.}
  \pline[1]{$A\leftarrow\varnothing$.}
  \pline[1]{\textbf{for each} shell $G(C,D)$ in this order \textbf{do}}
  \pline[2]{\textbf{if} $\mathcal L=\varnothing$ \textbf{then break}.}
  \pline[2]{Call $\textsc{ShellSearch}(C,D;\mathcal L,A)$.}
  \pline[1]{\textbf{end for}}
  \pline[1]{$\mathcal L\leftarrow P(R)\cap\Omega'$; $F\leftarrow$ the free vertices in $\bigcup\mathcal L$.}
  \pline{\textbf{end while}}
  \pline{\textbf{if} $\mathcal L=\varnothing$ \textbf{then return}.}
  \pline{$\omega\leftarrow$ the unique vertex in $F$.}
  \pline{$h\leftarrow\tfrac12\sum_{B\in\mathcal L}z(B)$.}
  \pline{\textbf{for each} $B\in\mathcal L$ \textbf{do}}
  \pline[1]{$\operatorname{Liquidate}(B)$.}
  \pline{\textbf{end for}}
  \pline{Run a Criterion-1 search from $\{\omega\}$ for exactly $h$ dual units.}
  \pline{$\mathcal L\leftarrow\varnothing$; \textbf{return}.}
\end{pseudocode}
\Description{Pseudocode of DismantlePath: while at least two free vertices remain, search the atomic shells in decreasing order of free vertices; then liquidate any remaining blossoms and run a final Criterion-1 search from the unique free vertex.}
\end{figure}

\textsc{ShellSearch} keeps its original boundaries $(C,D)$ but updates the current boundaries as inherited blossoms dissolve: at each iteration, $C^*$ is the smallest member of $\mathcal L$ containing $C$, and $D^*$ the largest member contained in $D$, or $\varnothing$ if there is none. The current shell is $G(C^*,D^*)$. If no undissolved ancestor of $C$ remains, the call ends. Otherwise it searches from the set $F^*$ of free vertices of $G(C^*,D^*)$ using only tight edges with both endpoints in $C^*\setminus D^*$, and contracts the maximal current blossoms that lie in $C^*\setminus D^*$; tight edges that cross a shell boundary are excluded.

Augmentation produces a maximum-cardinality matching in this contracted eligible subgraph by repeatedly flipping lifted augmenting paths, and it changes no matching edge outside $G(C^*,D^*)$. If no augmentation is possible, the search grows the alternating forest and contracts the odd cycles it reaches, giving each new current blossom dual zero. One unit of the ordinary dual adjustment then applies
\begin{align}
 y(u)&\leftarrow y(u)-\mathbf1_{u\in S}+\mathbf1_{u\in T}
       &&(u\in C^*\setminus D^*),\\
 z(B)&\leftarrow z(B)+2\cdot\mathbf1_{B\text{ outer}}-2\cdot\mathbf1_{B\text{ inner}}
       &&(B\text{ a maximal current blossom}),
 \label{eq:shell-current-update}
\end{align}
where $S$ and $T$ are the original vertices in outer and inner nodes of the shell search, a maximal current blossom is outer or inner according to its node, and $\mathbf1_P$ is one when $P$ holds and zero otherwise. This adjustment is paired with $\operatorname{Translate}(C^*)$ and, when $D^*\neq\varnothing$, with $\operatorname{Translate}(D^*)$; a translation decreases $z(B)$ by two and increases $y(u)$ by one for every $u\in B$. The vertices inside $D^*$ thus gain two and the other vertices of $C^*$ gain one, as in Appendix~\ref{app:scaling-shells}.

The call ends with the status \textsc{Augment} after an augmentation, with \textsc{Absorb} upon reaching vertices searched by an earlier call, and with \textsc{Dissolve} when the outermost member of $\mathcal L$ dissolves. Let $J$ be the set of vertices of the shells visited by this call. Every return invokes $\operatorname{Finish}(q)$, which sets $A\leftarrow A\cup J$ and returns the status $q$, so that absorption depends on earlier calls rather than on the shells that the present call has visited.

\begin{figure}[tbp]
\begin{pseudocode}{Algorithm 5}{ShellSearch$(C,D;\mathcal L,A)$}
  \pseudometa{Input}{An atomic inherited shell $G(C,D)$, the chain $\mathcal L$ and the earlier-search set $A$.}
  \pseudometa{Output}{Updated matching, duals and blossoms; the status \textsc{Augment}, \textsc{Absorb} or \textsc{Dissolve}.}
  \pline{$J\leftarrow\varnothing$.}
  \pline{\textbf{repeat}}
  \pline[1]{Resolve the current boundaries $C^*,D^*$ of $G(C,D)$.}
  \pline[1]{\textbf{if} $C^*$ does not exist \textbf{then return} $\operatorname{Finish}(\textsc{Dissolve})$.}
  \pline[1]{$F^*\leftarrow$ the free vertices of $G(C^*,D^*)$.}
  \pline[1]{\textbf{if} $(C^*\setminus D^*)\cap A\neq\varnothing$ \textbf{then return} $\operatorname{Finish}(\textsc{Absorb})$.}
  \pline[1]{$J\leftarrow J\cup(C^*\setminus D^*)$; $k\leftarrow|M|$.}
  \pline[1]{Augment to maximum cardinality using the tight eligible subgraph of $G(C^*,D^*)$.}
  \pline[1]{\textbf{if} $|M|>k$ \textbf{then return} $\operatorname{Finish}(\textsc{Augment})$.}
  \pline[1]{Grow from $F^*$ and contract the reachable current blossoms.}
  \pline[1]{Apply one unit of the ordinary dual adjustment in $G(C^*,D^*)$.}
  \pline[1]{$\operatorname{Translate}(C^*)$; \textbf{if} $D^*\neq\varnothing$ \textbf{then} $\operatorname{Translate}(D^*)$.}
  \pline[1]{Expand the maximal current blossoms with zero dual until none remains.}
  \pline[1]{$B_{\mathrm{out}}\leftarrow$ the outermost member of $\mathcal L$.}
  \pline[1]{Remove $C^*$ and $D^*$ from $\Omega'$ and $\mathcal L$ if their duals are zero.}
  \pline[1]{\textbf{if} $B_{\mathrm{out}}$ was removed \textbf{then return} $\operatorname{Finish}(\textsc{Dissolve})$.}
  \pline{\textbf{forever}}
\end{pseudocode}
\Description{Pseudocode of ShellSearch: a loop that resolves the current shell, augments when possible, and otherwise grows, contracts, adjusts duals, translates the boundaries, expands current blossoms with zero dual and removes dissolved boundaries.}
\end{figure}

The next iteration resolves the enlarged shell after any inherited boundary has been removed, and current blossoms with zero dual are expanded before the boundaries change. The listings specify dual adjustments of one unit; the implementation processes intervals without events in one step, up to the next enabled event or boundary, and obtains the maximum-cardinality matching of a shell through repeated maximal-augmentation phases (Section~\ref{sec:scaling-implementation}).

\section{Dual Adjustments and Transitions of Edmonds and Gabow}
\label{app:shared-contract}

The dual adjustment $\operatorname{Dual}(\mathcal X,\delta)$ acts on original vertices and maximal labelled blossoms:
\begin{align}
 y(v)&\leftarrow\begin{cases}y(v)-\delta,&v\in S,\\y(v)+\delta,&v\in T,\\y(v),&\text{otherwise},\end{cases}\\
 z(B)&\leftarrow\begin{cases}z(B)+2\delta,&B\text{ is maximal outer},\\z(B)-2\delta,&B\text{ is maximal inner},\\z(B),&\text{otherwise}.\end{cases}
\end{align}
With $\delta=\min_i\delta_i$, the duals stay feasible until the event with the smallest delta value becomes enabled. An edge inside a maximal labelled blossom keeps its dual value, because the change in $z(B)$ offsets the changes in the vertex duals.

The witness $\xi$ is a vertex for event~1, an edge for events~2 and~3, and a blossom for event~4. On the adjusted state, the enabled transition is
\begin{equation}
\mathcal X^+=\begin{cases}
 \operatorname{Settle}(\mathcal X,v),&i=1,\ \xi=v,\ y(v)=0,\\
 \operatorname{Grow}(\mathcal X,uv),&i=2,\ \xi=uv,\ B_v\text{ is matched},\\
 \operatorname{Augment}(\mathcal X,Q_{uv}),&i=2,\ \xi=uv,\ B_v\text{ is free},\\
 \operatorname{Augment}(\mathcal X,Q_{uv}),&i=3,\ \xi=uv,\ \operatorname{root}(B_u)\neq\operatorname{root}(B_v),\\
 \operatorname{Contract}(\mathcal X,C_{uv}),&i=3,\ \xi=uv,\ \operatorname{root}(B_u)=\operatorname{root}(B_v),\\
 \operatorname{Expand}(\mathcal X,B),&i=4,\ \xi=B,\ z(B)=0.
\end{cases}
\label{eq:classical-transition}
\end{equation}
Here $uv$ is a tight edge between different maximal blossoms with the endpoint labels of its event, $Q_{uv}$ is the corresponding augmenting path lifted to the original graph, and $C_{uv}$ is the fundamental odd cycle of $uv$ in $\mathcal A$. A grow step labels the matched target blossom inner and its mate outer. A contraction turns $C_{uv}$ into a new outer blossom $B$, adds $B$ to $\Omega$ with $z(B)=0$ and contracts it in $\mathcal A$. An expansion restores the subblossoms of an inner blossom whose dual is zero.

Growth, contraction and expansion preserve the matching, whereas augmentation flips $Q_{uv}$:
\begin{equation}
 M\leftarrow M\oplus Q_{uv},\qquad |M_{\mathrm{new}}|=|M_{\mathrm{old}}|+1.
\end{equation}
Settlement instead flips the even alternating path from the active root to $v$, which leaves $v$ free, and retires that tree. Trees that the settlement does not touch stay active, so one settlement does not end the algorithm. Per-tree settlement is an adaptation: the programs initialize each vertex dual to half the largest incident weight, whereas Galil, Micali and Gabow use half the global maximum and stop at their $\delta_1$ event~\cite{GalilMicaliGabow1986}.

Throughout the search, $M$ is a matching, $\Omega$ is laminar and every blossom in it is full, the duals are feasible ($y\geq0$ and $\pi(e)\geq0$ for every edge), and matched edges and the edges of every $E_B$ are tight. Fullness means $|M\cap E(B)|=\lfloor|B|/2\rfloor$ for every $B\in\Omega$. Once every free vertex has dual zero, matched-edge tightness and fullness give
\begin{equation}
 w(M)=\sum_{v\in V}y(v)+\sum_{B\in\Omega}\lfloor|B|/2\rfloor\,z(B).
\end{equation}
The right-hand side is the objective of a feasible dual solution, so by weak duality $M$ is a maximum weighted matching~\cite{GalilMicaliGabow1986}.

\section{Measurement Details}
\label{app:campaign-evidence}

All measurements except the counting, profiling and NCA-timing runs of Section~\ref{sec:instrumented} come from one campaign, with one binary that reads the SuiteSparse and generated files of Suite~A with a reader that accepts their formats. The programs are compiled with GCC~11.4.1 as C++17 with \texttt{-O3 -DNDEBUG -march=x86-64-v3 -mtune=generic -fopenmp}, with profiling and diagnostic output disabled. \Scal{} is built with the persistent event and grow frontier and the stored shell translations of Section~\ref{sec:scaling-implementation}, without the split-findmin grow driver.
\begin{itemize}
\item Nodes: exclusive allocations of 240\,GiB on nodes with two AMD EPYC~9334 processors (32 cores each, one thread per core, two NUMA nodes), Linux~5.14 and glibc~2.34; each process runs on one CPU with one OpenMP thread. Two nodes of the cluster that had shown file-system faults in earlier jobs were excluded.
\item Limits: 15 hours per process, followed by termination after a grace period of 60~s; 48 hours per allocation.
\item Order: Suite~A runs \Edm{} and \Gab{} back to back in an order that alternates across trials; Suite~B runs the three programs in an order that rotates across trials.
\item Trials: three per graph and program; six on \texttt{kmer\_V2a}, \texttt{europe\_osm}, \texttt{nlpkkt200}, \texttt{erb17}, \texttt{erb25} and \texttt{erb26}.
\item Memory: the peak resident set size of each process from \texttt{getrusage}, in GiB ($2^{30}$ bytes).
\item Validation: LEMON's \MWM{} solver on all 49 Suite~A graphs, and for Suite~B the LEMON optima of the admission screen.
\item Search comparison: one untimed instrumented pair of runs of \Edm{} and \Gab{} per graph, compared through step counts, step sequences and their matchings.
\item Instrumented runs (Section~\ref{sec:instrumented}): the same compiler, with \texttt{-g} added in the counting builds, in separate counting, profiling and NCA-timing builds; exclusive nodes of the same type, one CPU, 240\,GiB and a six-hour limit per process (16 or 64\,GiB and ten minutes for the NCA-timing runs).
\end{itemize}
Per-trial records, scripts and source code are in the repository given in Section~\ref{sec:implementation}.

\section{Phase Times and Operation Counts}
\label{app:profiles}

Tables~\ref{tab:profile-edmonds}--\ref{tab:profile-scaling} give the phase times behind Figure~\ref{fig:profiles}. A phase time is the median of the three profiled trials of a graph and program, and a total is the sum of the phase medians. For \Scal{}, Figure~\ref{fig:profiles} combines the preparation of the inherited blossoms with dismantling and shows all stages other than these and the Criterion-2 phases as other; none of those stages took more than 4.5\% of the time on any graph. Table~\ref{tab:counts} gives the operation counts of \Edm{} and \Gab{}, which are equal: the contractions; the vertices that they list, in millions, counting a vertex again for every blossom it enters; the largest contracted blossom; the deepest nesting; the expansions; and the nodes added to trees. The number of tree teardowns is $n$ on every graph, because each vertex starts its own tree and each tree is torn down once.

\begin{table}[!htbp]
\centering\footnotesize
\caption{Phase times of \Edm{} in seconds on the 21 instrumented graphs and, below the line, the 24 structured graphs: medians of three trials, phase by phase, and their sum. Selection is the time spent finding the next candidate, and other covers the remaining phases, including initialization, the computation of $\delta_1$, $\delta_2$ and $\delta_4$, and the dual adjustments.}
\label{tab:profile-edmonds}
\setlength{\tabcolsep}{2.2pt}
\begin{tabular}{@{}l r r r r r r r@{}}
\toprule\textbf{Graph} & \textbf{Total} & \textbf{Select.} & \textbf{Grow} & \textbf{Contr.} & \textbf{Exp.} & \textbf{Teardown} & \textbf{Other}\\\midrule
ER 250k d32 W1k (2) & 196.0 & 4.4 & 141.3 & 2.7 & 3.3 & 7.0 & 37.3\\
ER 250k d32 W1k (3) & 304.6 & 4.5 & 151.7 & 80.6 & 3.0 & 7.3 & 57.5\\
ER 500k d32 W1k (1) & 620.9 & 10.0 & 348.5 & 119.0 & 5.5 & 16.6 & 121.2\\
ER 500k d32 W1M (1) & 462.6 & 9.9 & 327.7 & 1.0 & 9.7 & 15.4 & 98.9\\
ER 500k d32 W1k (2) & 502.3 & 10.1 & 356.5 & 4.6 & 6.4 & 16.7 & 108.1\\
ER 500k d64 W1k (2) & 1546.0 & 21.6 & 1249.5 & 15.0 & 14.8 & 29.1 & 216.0\\
ER 750k d32 W1k (2) & 850.4 & 16.1 & 613.1 & 3.6 & 10.2 & 26.4 & 181.2\\
ER 750k d32 W1k (3) & 3074.4 & 19.1 & 679.2 & 1827.8 & 10.6 & 27.9 & 509.7\\
ER 1024k d32 W1k (1) & 2088.7 & 25.0 & 1153.7 & 414.5 & 32.3 & 43.6 & 419.7\\
ER 1024k d32 W1k (2) & 1381.1 & 23.0 & 994.8 & 9.6 & 17.0 & 39.6 & 297.1\\
Bip. ER 500k d32 W1k (1) & 472.7 & 9.7 & 377.4 & 0.0 & 0.0 & 21.4 & 64.1\\
SBM 500k 28/4 W1k (2) & 471.5 & 10.0 & 343.3 & 1.6 & 6.0 & 15.1 & 95.6\\
WS 500k d32 b0.05 W1k (1) & 121.9 & 6.1 & 44.9 & 35.5 & 1.2 & 7.3 & 27.1\\
WS 500k d32 b0.05 W1k (3) & 72.8 & 5.1 & 39.8 & 0.6 & 1.0 & 6.6 & 19.6\\
Torus 500k d32 W1k (3) & 69.9 & 6.2 & 33.4 & 1.8 & 1.5 & 8.9 & 18.1\\
Clique 512k K32 W1k (1) & 15.8 & 3.1 & 5.9 & 0.2 & 0.1 & 2.4 & 4.2\\
Biclique 1024k K32,32 W1k (1) & 35.8 & 6.7 & 14.4 & 0.0 & 0.0 & 6.0 & 8.7\\
spal\_004 & 64.1 & 36.2 & 9.6 & 0.0 & 0.0 & 8.4 & 10.0\\
erb16 & 565.4 & 134.6 & 308.4 & 0.0 & 0.0 & 68.9 & 53.5\\
rgg\_n\_2\_23\_s0 & 568.5 & 107.6 & 205.5 & 10.4 & 11.0 & 86.0 & 148.0\\
kron\_g500-logn21 & 1044.7 & 110.6 & 733.6 & 2.5 & 0.0 & 150.8 & 47.1\\
\midrule
Biclique blocks (1) & 157.7 & 124.6 & 4.6 & 0.0 & 0.0 & 14.5 & 13.9\\
Biclique blocks (2) & 157.0 & 124.4 & 4.7 & 0.0 & 0.0 & 14.5 & 13.5\\
Connected blocks (1) & 157.2 & 124.4 & 4.6 & 0.0 & 0.0 & 14.4 & 13.8\\
Connected blocks (2) & 156.5 & 124.0 & 4.6 & 0.0 & 0.0 & 14.4 & 13.4\\
Biclique ring (1) & 158.4 & 124.9 & 5.2 & 0.0 & 0.0 & 14.5 & 13.8\\
Biclique ring (2) & 157.2 & 124.1 & 5.1 & 0.0 & 0.0 & 14.5 & 13.6\\
Ferrers blocks (1) & 133.6 & 104.6 & 7.3 & 0.0 & 0.0 & 10.9 & 10.7\\
Ferrers blocks (2) & 133.0 & 103.9 & 7.2 & 0.0 & 0.0 & 10.9 & 10.9\\
Convex windows (1) & 78.1 & 56.2 & 7.3 & 0.0 & 0.0 & 7.2 & 7.5\\
Convex windows (2) & 78.5 & 56.6 & 7.2 & 0.0 & 0.0 & 7.1 & 7.5\\
Circular intervals (1) & 144.8 & 110.2 & 7.1 & 0.0 & 0.0 & 13.9 & 13.5\\
Circular intervals (2) & 144.9 & 110.0 & 7.2 & 0.0 & 0.0 & 13.9 & 13.8\\
Cube neighbourhoods (1) & 131.6 & 90.1 & 18.2 & 0.0 & 0.0 & 11.5 & 11.7\\
Cube neighbourhoods (2) & 156.4 & 114.7 & 18.4 & 0.0 & 0.0 & 11.6 & 11.7\\
Odd cliques (1) & 226.1 & 103.2 & 80.1 & 3.2 & 1.5 & 13.5 & 24.6\\
Odd cliques (2) & 227.4 & 104.2 & 80.3 & 3.0 & 1.6 & 13.5 & 24.8\\
Clique chain (1) & 228.6 & 102.7 & 80.1 & 3.4 & 1.9 & 13.8 & 26.7\\
Clique chain (2) & 228.8 & 103.4 & 79.6 & 3.3 & 2.0 & 13.8 & 26.8\\
Tripartite (1) & 176.0 & 68.6 & 69.0 & 3.0 & 1.4 & 12.1 & 22.0\\
Tripartite (2) & 173.8 & 67.9 & 68.6 & 2.9 & 1.3 & 11.6 & 21.5\\
Five-cycle blow-ups (1) & 232.5 & 83.3 & 102.6 & 3.4 & 1.7 & 14.5 & 26.9\\
Five-cycle blow-ups (2) & 233.4 & 83.5 & 103.2 & 3.4 & 1.8 & 14.6 & 26.9\\
Cycle powers (1) & 140.0 & 50.4 & 56.2 & 3.0 & 1.2 & 10.1 & 19.2\\
Cycle powers (2) & 140.0 & 50.2 & 56.3 & 2.9 & 1.2 & 10.1 & 19.2\\
\bottomrule\end{tabular}\end{table}

\begin{table}[!htbp]
\centering\footnotesize
\caption{Phase times of \Gab{} in seconds on the graphs and with the phases of Table~\ref{tab:profile-edmonds}.}
\label{tab:profile-gabow}
\setlength{\tabcolsep}{2.2pt}
\begin{tabular}{@{}l r r r r r r r@{}}
\toprule\textbf{Graph} & \textbf{Total} & \textbf{Select.} & \textbf{Grow} & \textbf{Contr.} & \textbf{Exp.} & \textbf{Teardown} & \textbf{Other}\\\midrule
ER 250k d32 W1k (2) & 206.3 & 0.2 & 144.5 & 3.2 & 3.3 & 14.7 & 40.5\\
ER 250k d32 W1k (3) & 356.5 & 0.3 & 156.2 & 132.2 & 3.0 & 15.2 & 49.6\\
ER 500k d32 W1k (1) & 710.8 & 0.6 & 356.5 & 195.9 & 5.4 & 33.4 & 119.2\\
ER 500k d32 W1M (1) & 480.6 & 0.5 & 333.8 & 1.6 & 9.3 & 31.3 & 104.1\\
ER 500k d32 W1k (2) & 522.7 & 0.5 & 362.6 & 6.0 & 6.4 & 33.3 & 114.0\\
ER 500k d64 W1k (2) & 1591.4 & 0.8 & 1261.2 & 22.9 & 18.3 & 61.3 & 226.9\\
ER 750k d32 W1k (2) & 881.3 & 0.8 & 623.4 & 4.6 & 10.4 & 51.9 & 190.1\\
ER 750k d32 W1k (3) & 3520.6 & 1.1 & 695.3 & 2474.0 & 10.4 & 54.4 & 285.4\\
ER 1024k d32 W1k (1) & 2500.2 & 1.4 & 1181.7 & 611.7 & 207.3 & 82.9 & 415.1\\
ER 1024k d32 W1k (2) & 1428.6 & 1.2 & 1008.8 & 14.7 & 17.5 & 76.1 & 310.3\\
Bip. ER 500k d32 W1k (1) & 490.6 & 0.5 & 384.8 & 0.0 & 0.0 & 35.6 & 69.7\\
SBM 500k 28/4 W1k (2) & 486.6 & 0.4 & 349.1 & 2.1 & 6.0 & 29.2 & 99.8\\
WS 500k d32 b0.05 W1k (1) & 152.8 & 0.3 & 48.7 & 59.8 & 1.4 & 15.1 & 27.6\\
WS 500k d32 b0.05 W1k (3) & 81.2 & 0.3 & 42.8 & 1.1 & 1.1 & 13.8 & 22.1\\
Torus 500k d32 W1k (3) & 86.2 & 0.4 & 37.2 & 3.4 & 1.9 & 20.1 & 23.3\\
Clique 512k K32 W1k (1) & 20.2 & 0.2 & 6.9 & 0.7 & 0.1 & 6.2 & 6.0\\
Biclique 1024k K32,32 W1k (1) & 43.4 & 0.4 & 16.8 & 0.0 & 0.0 & 14.0 & 12.2\\
spal\_004 & 62.8 & 0.0 & 11.1 & 0.0 & 0.0 & 30.9 & 20.7\\
erb16 & 597.4 & 0.2 & 315.9 & 0.0 & 0.0 & 169.5 & 111.8\\
rgg\_n\_2\_23\_s0 & 621.2 & 5.1 & 241.5 & 23.9 & 11.2 & 155.2 & 184.4\\
kron\_g500-logn21 & 1164.1 & 0.8 & 764.3 & 0.2 & 0.0 & 290.3 & 108.5\\
\midrule
Biclique blocks (1) & 74.9 & 0.0 & 5.1 & 0.0 & 0.0 & 45.4 & 24.4\\
Biclique blocks (2) & 74.3 & 0.0 & 5.2 & 0.0 & 0.0 & 45.3 & 23.7\\
Connected blocks (1) & 79.4 & 0.0 & 5.1 & 0.0 & 0.0 & 47.4 & 26.9\\
Connected blocks (2) & 79.5 & 0.0 & 5.2 & 0.0 & 0.0 & 47.7 & 26.5\\
Biclique ring (1) & 75.7 & 0.0 & 5.9 & 0.0 & 0.0 & 45.7 & 24.1\\
Biclique ring (2) & 75.4 & 0.0 & 5.9 & 0.0 & 0.0 & 45.4 & 24.0\\
Ferrers blocks (1) & 69.9 & 0.0 & 8.2 & 0.0 & 0.0 & 42.3 & 19.4\\
Ferrers blocks (2) & 70.2 & 0.0 & 8.1 & 0.0 & 0.0 & 42.1 & 20.0\\
Convex windows (1) & 45.0 & 0.0 & 7.6 & 0.0 & 0.0 & 24.9 & 12.4\\
Convex windows (2) & 44.9 & 0.0 & 7.6 & 0.0 & 0.0 & 25.0 & 12.2\\
Circular intervals (1) & 89.7 & 0.0 & 8.5 & 0.0 & 0.0 & 57.4 & 23.7\\
Circular intervals (2) & 89.9 & 0.0 & 8.6 & 0.0 & 0.0 & 57.4 & 23.8\\
Cube neighbourhoods (1) & 92.4 & 0.0 & 22.1 & 0.0 & 0.0 & 48.3 & 22.0\\
Cube neighbourhoods (2) & 92.1 & 0.0 & 22.0 & 0.0 & 0.0 & 48.1 & 22.1\\
Odd cliques (1) & 199.2 & 0.6 & 93.7 & 22.0 & 1.7 & 44.6 & 36.5\\
Odd cliques (2) & 200.5 & 0.6 & 95.0 & 22.0 & 1.8 & 44.5 & 36.5\\
Clique chain (1) & 200.2 & 0.6 & 94.0 & 21.3 & 1.9 & 44.9 & 37.5\\
Clique chain (2) & 199.9 & 0.6 & 93.4 & 21.3 & 1.9 & 45.0 & 37.6\\
Tripartite (1) & 167.0 & 0.7 & 81.7 & 17.0 & 1.5 & 34.0 & 32.0\\
Tripartite (2) & 164.9 & 0.7 & 80.6 & 16.9 & 1.5 & 33.2 & 32.0\\
Five-cycle blow-ups (1) & 218.7 & 0.7 & 117.3 & 18.2 & 1.9 & 41.8 & 38.8\\
Five-cycle blow-ups (2) & 219.3 & 0.7 & 117.6 & 18.2 & 1.9 & 42.0 & 38.9\\
Cycle powers (1) & 135.0 & 0.7 & 65.2 & 14.3 & 1.3 & 26.9 & 26.5\\
Cycle powers (2) & 135.3 & 0.7 & 65.4 & 14.3 & 1.3 & 27.0 & 26.6\\
\bottomrule\end{tabular}\end{table}

\begin{table}[!htbp]
\centering\footnotesize
\caption{Stage times of \Scal{} on the instrumented Suite~B graphs in seconds: medians of three trials, stage by stage, summed over the scales. Rescale reveals the next weight bit and updates the weights and duals; Prep.\ prepares the inherited blossoms, including liquidation; Dism.\ is dismantling; FVR~p.\ prepares free-vertex reduction; Crit.~2 and Crit.~3 are the Criterion-2 and Criterion-3 phases of FVR, and Perf.\ is Perfection; other covers setup, carrying the weights to the next scale, finalization and output.}
\label{tab:profile-scaling}
\setlength{\tabcolsep}{1.6pt}
\begin{tabular}{@{}l r r r r r r r r r@{}}
\toprule\textbf{Graph} & \textbf{Total} & \textbf{Rescale} & \textbf{Prep.} & \textbf{Dism.} & \textbf{FVR p.} & \textbf{Crit.~2} & \textbf{Crit.~3} & \textbf{Perf.} & \textbf{Other}\\\midrule
ER 250k d32 W1k (2) & 39.6 & 0.5 & 0.4 & 0.3 & 0.3 & 37.6 & 0.0 & 0.0 & 0.4\\
ER 250k d32 W1k (3) & 40.5 & 0.5 & 0.4 & 0.4 & 0.3 & 38.5 & 0.0 & 0.0 & 0.4\\
ER 500k d32 W1k (1) & 123.4 & 1.0 & 0.8 & 0.6 & 0.6 & 119.4 & 0.1 & 0.0 & 0.8\\
ER 500k d32 W1M (1) & 137.2 & 1.3 & 1.1 & 0.8 & 0.9 & 132.0 & 0.0 & 0.0 & 0.9\\
ER 500k d32 W1k (2) & 123.5 & 1.0 & 0.8 & 0.7 & 0.7 & 119.4 & 0.1 & 0.0 & 0.8\\
ER 500k d64 W1k (2) & 207.2 & 2.0 & 2.1 & 0.7 & 2.0 & 198.9 & 0.1 & 0.0 & 1.5\\
ER 750k d32 W1k (2) & 288.1 & 1.6 & 1.2 & 1.0 & 1.0 & 282.0 & 0.1 & 0.0 & 1.2\\
ER 750k d32 W1k (3) & 276.6 & 1.6 & 1.3 & 1.0 & 1.0 & 270.4 & 0.0 & 0.0 & 1.2\\
ER 1024k d32 W1k (1) & 455.9 & 2.2 & 1.7 & 1.3 & 1.3 & 447.7 & 0.0 & 0.0 & 1.7\\
ER 1024k d32 W1k (2) & 495.2 & 2.2 & 1.7 & 1.5 & 1.4 & 486.4 & 0.4 & 0.0 & 1.7\\
Bip. ER 500k d32 W1k (1) & 105.6 & 1.0 & 0.8 & 0.0 & 0.6 & 102.3 & 0.0 & 0.0 & 0.8\\
SBM 500k 28/4 W1k (2) & 93.4 & 1.0 & 0.8 & 0.7 & 0.6 & 89.3 & 0.1 & 0.0 & 0.8\\
WS 500k d32 b0.05 W1k (1) & 85.6 & 1.0 & 1.1 & 7.2 & 0.7 & 74.7 & 0.0 & 0.0 & 0.8\\
WS 500k d32 b0.05 W1k (3) & 86.3 & 1.0 & 1.1 & 7.5 & 0.7 & 75.0 & 0.0 & 0.0 & 0.8\\
Torus 500k d32 W1k (3) & 2347.4 & 1.1 & 84.2 & 1044.3 & 0.7 & 1215.4 & 0.9 & 0.0 & 0.8\\
Clique 512k K32 W1k (1) & 27.4 & 1.0 & 1.0 & 3.5 & 0.6 & 20.4 & 0.0 & 0.0 & 0.8\\
Biclique 1024k K32,32 W1k (1) & 48.0 & 2.2 & 1.7 & 0.0 & 1.4 & 41.0 & 0.0 & 0.0 & 1.6\\
\bottomrule\end{tabular}\end{table}

\begin{table}[!htbp]
\centering\footnotesize
\caption{Operation counts of \Edm{} and \Gab{} on the instrumented graphs.}
\label{tab:counts}
\setlength{\tabcolsep}{2.8pt}
\begin{tabular}{@{}l r r r r r r@{}}
\toprule\textbf{Graph} & \textbf{Contr.} & \textbf{Listed (M)} & \textbf{Largest} & \textbf{Nesting} & \textbf{Exp.} & \textbf{Tree adds}\\\midrule
ER 250k d32 W1k (2) & 4\,829 & 161.2 & 56\,631 & 4\,218 & 608 & 1\,908\,309\\
ER 250k d32 W1k (3) & 43\,228 & 5\,126.2 & 188\,113 & 41\,532 & 1\,691 & 2\,128\,125\\
ER 500k d32 W1k (1) & 35\,039 & 5\,235.0 & 234\,087 & 33\,752 & 1\,280 & 4\,156\,197\\
ER 500k d32 W1M (1) & 2\,338 & 32.3 & 29\,845 & 557 & 1\,775 & 3\,766\,673\\
ER 500k d32 W1k (2) & 7\,716 & 272.3 & 73\,063 & 3\,535 & 4\,175 & 4\,157\,185\\
ER 500k d64 W1k (2) & 15\,944 & 966.3 & 121\,619 & 9\,578 & 13\,134 & 4\,325\,764\\
ER 750k d32 W1k (2) & 5\,414 & 174.4 & 67\,609 & 2\,052 & 4\,876 & 6\,229\,480\\
ER 750k d32 W1k (3) & 130\,247 & 47\,653.3 & 571\,427 & 127\,440 & 2\,800 & 7\,148\,499\\
ER 1024k d32 W1k (1) & 62\,227 & 12\,660.0 & 379\,181 & 43\,919 & 51\,358 & 10\,617\,601\\
ER 1024k d32 W1k (2) & 10\,588 & 558.6 & 106\,745 & 3\,625 & 6\,959 & 9\,220\,249\\
Bip. ER 500k d32 W1k (1) & 0 & 0.0 & 0 & 0 & 0 & 5\,116\,016\\
SBM 500k 28/4 W1k (2) & 3\,678 & 67.7 & 46\,259 & 1\,362 & 2\,238 & 4\,097\,880\\
WS 500k d32 b0.05 W1k (1) & 46\,340 & 2\,305.1 & 185\,913 & 18\,267 & 12\,778 & 2\,722\,030\\
WS 500k d32 b0.05 W1k (3) & 26\,063 & 23.3 & 36\,771 & 745 & 10\,755 & 2\,373\,932\\
Torus 500k d32 W1k (3) & 39\,834 & 109.6 & 64\,053 & 2\,544 & 21\,102 & 2\,066\,270\\
Clique 512k K32 W1k (1) & 29\,585 & 0.3 & 31 & 11 & 3\,250 & 661\,721\\
Biclique 1024k K32,32 W1k (1) & 0 & 0.0 & 0 & 0 & 0 & 1\,416\,908\\
spal\_004 & 0 & 0.0 & 0 & 0 & 0 & 21\,556\\
erb16 & 0 & 0.0 & 0 & 0 & 0 & 349\,126\\
rgg\_n\_2\_23\_s0 & 894\,676 & 32.5 & 2\,369 & 288 & 291\,014 & 22\,779\,894\\
kron\_g500-logn21 & 3 & 0.0 & 7 & 1 & 3 & 2\,513\,148\\
\bottomrule\end{tabular}\end{table}

\end{document}